\documentclass[5p,twocolumn,sort&compress]{elsarticle}

\usepackage{commath}
\usepackage{amsmath}
\usepackage{amssymb}
\usepackage{color}
\usepackage{url}
\usepackage{graphicx}
\usepackage{bm}
\usepackage{hyperref}
\usepackage{multicol}
\usepackage{braket}

\hypersetup{colorlinks=true,citecolor=blue, linkcolor=blue, urlcolor=blue}

\begin{document}

\begin{frontmatter}

\title{Revisiting a non-parametric reconstruction of the deceleration parameter from combined background and the growth rate data}

\author{Purba Mukherjee\corref{pm}%
\cortext[pm]{Corresponding Author} \fnref{myfootnote}}

\author{Narayan Banerjee\fnref{myfootnote2}}

\address{$^{1,2}$Department of Physical Sciences,  \\Indian Institute of Science Education and Research Kolkata,\\ Mohanpur, West Bengal 741246, India.}

\fntext[myfootnote]{Email: pm14ip011@iiserkol.ac.in (Purba Mukherjee)}
\fntext[myfootnote2]{Email: narayan@iiserkol.ac.in (Narayan Banerjee)}



\begin{abstract}

The cosmic deceleration parameter $q$ has been reconstructed in a non-parametric way using various combinations of recent observational datasets. The Pantheon 
compilation of the Supernova (SN) distance modulus data, the Cosmic Chronometer (CC) measurements of the Hubble parameter including the full systematics and 
the Baryon Acoustic Oscillation (BAO) data have been considered in this work. The redshift $z_t$, where the transition from a past decelerated to a late-time 
accelerated phase of evolution occurs, is estimated from the reconstructed $q$. The possible effect of a non-zero spatial curvature from the Planck 2020 estimate 
is checked. The outcome of including different $H_0$ measurements from recent Planck 2020 and Riess 2021 probes having a maximum discrepancy at the $4.2\sigma$ 
level, is investigated. Results indicate that the transition from a past decelerated phase to the late-time accelerated phase occurs within the redshift range 
$0.5<z<1$. For $z>1$, the reconstructed $q$ is observed to have a non-monotonic evolution in case of the combined CC and SN data. On introducing the BAO data, 
the reconstructed $q$ shows an oscillating behaviour for $z\gtrsim1$. To investigate the effect of matter perturbations, the growth rate data from the Redshift-Space 
Distortions (RSD) are utilized in reconstructing $q$. Using the $\mathcal{O}m(z)$ diagnostic, we draw inferences on the validity of $\Lambda$CDM as a consistency 
check. The $\Lambda$CDM model is well consistent and included at the 2$\sigma$ level in the domain of all the reconstructions.  

\end{abstract}


\end{frontmatter}


\section{Introduction}

Although observations have established a recent accelerated expansion of the universe \cite{riess, perl}, the nature of the agent driving this acceleration is yet 
to be established. For a comprehensive study on the diverse aspects of this accelerated expansion and various tensions between observations, we refer to \cite{diego, 
zzhai,joan_adria,rafael,valentino,cosmoletter1,cosmoletter2,cosmoletter3} and references therein. A variety of theoretical models proposed can explain this accelerated 
expansion, either in the form of an additional field called dark energy in the matter sector or in the form of modifying the theory of gravity itself, yet all of them 
have the generic problem of not being desperately required by any other branch of physics \cite{brax}. 

This inspires a reverse way of looking at the evolution. Rather than trying to find the evolution from the given matter sector using Einstein field equations, one 
uses the evolutionary history that fits with observations, to find out the possible distribution of matter. Normally physical quantities like the equation of state 
parameter of the dark energy \cite{saini, varun}, the quintessence potential \cite{staro, huterer1, huterer2} occupy the central stage of interest in this game of 
reconstruction. A recent trend of reconstruction ignores any dynamical equation and makes an attempt towards finding out the kinematical quantities, that are defined 
as the time derivatives of the scale factor $a(t)$, directly from observations. Being the first order derivative of the scale factor, the Hubble parameter $H$ that 
measures the rate of cosmic expansion is found to evolve with time. So, the natural choice as the relevant parameters are the next higher order derivatives like the 
deceleration parameter $q$ and the jerk parameter $j$. It should be mentioned that these cosmological parameters also suffer from tensions between different datasets, 
which is particularly true for the measurement of the present value of the Hubble parameter $H_0$ \cite{diego,dhawan,cosmoletter2,planck2020,wendy,riess2021,aiola,
macaulay}. 

Already there are quite a few investigations in this direction. Reconstruction of the deceleration parameter $q$ naturally started quite a long time back and can be 
found in the work of Gong and Wang \cite{gong1, gong2}, Wang, Xu, Lu and Gui \cite{wang}, Lobo, Mimoso and Visser \cite{jose}, Mamon and Das \cite{sudipta}, Mamon 
\cite{mamon}, Cardenas and Motta \cite{motta}, Jesus, Holanda and Pereira \cite{jesus}, Yang and Gong \cite{yangong}, Almada et al. \cite{almada2020}. The list is 
surely not quite exhaustive. As $q$ is evolving, the next higher order derivative, the jerk parameter $j$ has also been reconstructed from observational data 
\cite{luongo, rapetti, zz, ankan1, ankan2}. These investigations mostly rely on a parametrization of the kinematical quantity and an estimation of the parameters 
from observational data. This approach normally is a bit biased as the quantities depend on $z$ in a given way according to the functional form chosen as an ansatz. 
A more robust form is a non-parametric reconstruction, where the quantity of interest is reconstructed directly from the data without assuming any functional form. 
For the physical quantities like the equation of state parameter of the dark energy, dark energy potential, etc., this practice is already there \cite{sahl1, sahl2, 
holsclaw1, holsclaw2, holsclaw3, critt, sanjay, zzhang}. 

The motivation of this work is to reconstruct  the deceleration parameter $q$ directly from observational data without assuming any parametric form for $q$. We do 
not start from any theory of gravity or any form of matter distribution in the universe. The only a priori assumption is that the universe is spatially homogeneous 
and isotropic, thus described by the Friedmann-Lemaitre-Robertson-Walker (FLRW) metric. We restrict ourselves to the reconstruction of the kinematical parameter $q$ 
and do not aim to figure out physical quantities like the dark energy potential or the dark energy equation of state. 

There are already examples of a non-parametric reconstruction of $q$ in literature. Bilicki and Seikel \cite{bilicki} reconstructed $q$ using Union 2.1 \cite{union2.1} 
compilation for the Supernova data. Lin, Li and Tang \cite{lin} did a similar reconstruction with the Pantheon \cite{pan1} compilation for the Supernova data with 
various priors for $H_0$. A slightly older similar work in \cite{xia} uses Union 2 \cite{union2} and Union 2.1 compilations with various priors for $H_0$. Recently, 
Nunes et al. \cite{nunes} reconstructed $q$ using the transversal Baryon Acoustic Oscillation data. Evidence for cosmic acceleration up to the 7$\sigma$ level with 
next-generation surveys like Euclid and SKA was obtained by Bengaly \cite{carlos}. Another non-parametric reconstruction of $q$ and an estimation of the transition 
redshift was done by Jesus, Valentim, Escobal and Pereira \cite{jesus_nonpara}. One can refer to the works of Arjona and Nesseris \cite{arjona}, Velten, Gomes and 
Busti \cite{busti}, G\'{o}mez-Valent \cite{adria} and Haridasu et al. \cite{haridasu} for more information on the non-parametric reconstruction of $q$. 

The aim of the present work is to revisiting the non-parametric reconstruction of $q$ using various combination of recent datasets, adopting the Gaussian Process 
\cite{william,mackay,rw} technique. In the absence of a universally accepted form of dark energy, this kind of revisit is an essential tool for refining the present 
understanding of the accelerated expansion of the universe. Our work is close to the investigations by \cite{bilicki,lin,xia,jesus_nonpara} and \cite{haridasu}, but 
with difference from each of them in the data sets and the methodology followed. Different combinations of the Supernova distance modulus compilation, Cosmic 
Chronometer Hubble parameter measurements and Baryon Acoustic Oscillations have been utilized for reconstructing $q$. The late-time transition redshift $z_t$, where 
the universe undergoes a transition from a decelerated to an accelerated phase of expansion, has been estimated. A prior choice on the $H_0$ value from Planck 2020 
\cite{planck2020} estimate and Riess 2021 \cite{riess2021} measurement with a 4.2$\sigma$ tension between them, has also been checked. The effect of spatial curvature 
which has mostly been ignored for simplicity in the previous works, except the work of Zhang and Xia \cite{xia}, has been studied. As the growth of perturbations play 
a promising role in distinguishing among diverse dark energy models, we further utilize the growth rate measurements from the Redshift-Space Distortions which has 
commonly been ignored for a reconstruction of $q$ in the previous works. We obtain the Hubble parameter at the present epoch $H_0$ for a combination of datasets in 
a novel way which serves as a normalization constant for the Hubble and SN comoving distance data in the final GP reconstruction. We also reconstruct the $\mathcal{O}
m(z)$ diagnostics \cite{om1, om2, om3} simultaneously as alternative investigators to detect possible deviations from $\Lambda$CDM, as studied in \cite{haridasu}. 

Results obtained in the present work clearly show that the standard $\Lambda$CDM model is well consistent at the 2$\sigma$ level in the domain of all the reconstructions. 
The reconstructed $q$ from the background data shows the possibility of a negative dip at higher redshift values. Thus, the deceleration preceding the present acceleration 
might have been a transient phenomena. However, this behaviour may be statistically not too significant as a positive $q$ is comfortably included at the $1\sigma$ confidence 
level. The use of any prior measurement for $H_0$, and the spatial curvature density parameter $\Omega_{k,0}$ does not make any qualitative difference in this regard. The 
matter density parameter $\Omega_{m,0}$ is observed to have a strong influence on the reconstruction from the growth rate data. We shall compare our method and the results 
obtained with those of the existing literature in the final section. This comparison can also be used as an inventory of results. 

The paper is organized as follows. The following section introduces the theoretical framework for the cosmological parameters that we shall be dealing with. In 
section 3, the observational datasets have been briefly reviewed. Section 4 describes the methodology adopted. Reconstruction using different combinations of the 
background datasets is presented in section 5. Reconstruction with the growth rate data is presented in section 6. We conclude the manuscript in section 7 with 
an overall discussion about the results.

\section{Theoretical framework}

A spatially homogeneous and isotropic universe is given by the Friedmann-Lema\^{i}tre-Robertson-Walker (FLRW) metric
\begin{equation}
\small \dif s^2 = -c^2 \dif t^2 + a^2(t) \left[\frac{\dif r^2}{1-kr^2} + r^2 \dif\theta^2 + r^2 \sin^2\theta \dif\phi^2 \right],  \label{flrw}
\end{equation} 

where $a(t)$ is the scale factor and $k (=0, \pm 1)$ is the curvature index.  \\

The Hubble parameter is defined as
\begin{equation}
H = \frac{\dot{a}}{a} ,
\end{equation}  

where a `dot' denotes derivative with respect to the cosmic time $t$.

All cosmological parameters can be rewritten as functions of the redshift $z$, defined as $1+z = \frac{a_0}{a}$. For convenience, we define the reduced 
Hubble parameter as

\begin{equation}
E(z) = \frac{H(z)}{H_0} ,
\end{equation} 

where a subscript $0$ indicates the present value of the corresponding quantity.

The transverse comoving distance $d_C$ of luminous objects, like supernovae, is given by

\begin{equation} \label{D}
d_C(z)=  \frac{c}{H_0 \sqrt{\vert \Omega_{k,0} \vert}}\sin\mbox{$n$} \left( \sqrt{\vert \Omega_{k,0} \vert } \int_{0}^{z} \frac{\dif z'}{E(z')}\right) ,
\end{equation} 

in which the $\sin n$ function is a shorthand for

\[ \sin nx= \begin{cases}
\sinh x &  (\Omega_{k,0}>0), \\
~~x & (\Omega_{k,0} = 0), \\
\sin x & (\Omega_{k,0}<0).
\end{cases}
\]

The dimensionless quantity $\Omega_{k,0} = -\frac{k c^2}{a_0^2 H_0^2}$, called the cosmic curvature density parameter is positive, negative or zero corresponding 
to the spatial curvature $k = -1, +1, 0$ which signifies an open, closed, or flat universe, respectively. Equation \eqref{D} can be represented in a dimensionless 
way, known as the normalised comoving distance, as

\begin{equation}
D(z) = \frac{H_0 d_C}{c}. \label{D_from_dc}
\end{equation}

The deceleration parameter is a dimensionless measure of the cosmic acceleration and is defined by

\begin{equation}\label{qdef}
q = -\frac{\ddot{a}}{a H^2}.
\end{equation} 

Cosmological observations indicate that the universe is undergoing an accelerated expansion in the recent epoch, i.e., $q < 0$. However, this acceleration must have 
set in during a recent past and is not a permanent feature of the evolution. This transition from a decelerated to an accelerated phase of expansion is marked by a 
change in signature of $q$, which occurs at some particular $z_t$, known as the deceleration-acceleration transition redshift.

\section{Observational Data} \label{obs-data}

In this work we use different combinations of datasets like the Cosmic Chronometer data (CC), the Type Ia Supernova (SN) distance modulus data and the Baryon 
Acoustic Oscillation data (BAO) for reconstructing the cosmic deceleration parameter $q$ as a function of the redshift $z$. In the beginning, we reconstruct $q$ 
from the combination of CC and SN data. As there are apprehensions that the BAO measurements in galaxy surveys depend crucially on a fiducial cosmological model, 
we also reconstruct $q$ from the combined CC, SN and BAO datasets, to examine the possible effect of BAO data on the reconstruction. The growth rate $f\sigma_8$ 
measurements from the redshift-space distortions (RSD), caused by the peculiar motions of galaxies \cite{kaiser}, are further taken into account for another 
reconstruction of $q$. A brief summary of the datasets is given below.

\subsection{CC Data}

The Hubble parameter $H(z)$ can directly be measured by calculating the differential ages of galaxies \cite{cc_101,cc_102,cc_103,cc_104,cc_106,cc_105}, known as 
the cosmic chronometers (CC), given by 

\begin{equation} \label{ccH}
H(z) = -\frac{1}{1+z} \frac{\dif z}{\dif t}.
\end{equation}

These measurements are independent of the Cepheid distance scale and do not rely on any particular cosmological model. But they are subject to other sources of 
systematic uncertainties, such as those associated with the modelling of stellar ages, which is carried out through the so-called stellar population synthesis 
(SPS) techniques. Given a pair of ensembles of passively evolving galaxies at two different redshift points, it is possible to infer $\frac{\dif z}{\dif t}$ 
from observations and under the assumption of a concrete SPS model \cite{cc_103, cc_104, sps2}. Therefore, one can obtain direct information about the Hubble function 
at different $z$. 
`
In the present work, we take into account the CCB compilation consisting of 31 points and the CCM compilation that consists of 15 points, obtained by considering 
the BC03 \cite{bc03} and MaStro \cite{mastro} SPS models, respectively. Systematic effects for the CC samples, broadly discussed in Moresco et al. \cite{sps2}, 
have been added to the covariance matrices of the current CC data for representing the full range of errors.

\subsection{SN-Ia Data}

For the supernova data, we use the recent Pantheon compilation by Scolnic et al. \cite{pan1}. The numerical data of the full Pantheon SN-Ia catalogue is publicly 
available\footnote{\url{http://dx.doi.org/10.17909/T95Q4X}}$^,$\footnote{\url{https://archive.stsci.edu/prepds/ps1cosmo/index.html}} with a detailed description. 
The Pantheon compilation is presently the largest spectroscopically confirmed SN-Ia sample, which consists of 1048 supernovae from different surveys, including 
the Sloan Digital Sky Survey (SDSS) \cite{kessler2009}, SN Legacy Survey (SNLS) \cite{conley2011}, various low-z samples viz. the Pan-STARRS1 Medium Deep Survey 
\cite{rest2014}, the Harvard Smithsonian Center for Astrophysics SN surveys \cite{hicken2009}, the Carnegie SN Project \cite{strit2011} and some high-z data from 
the Hubble Space Telescope (HST) cluster SN survey \cite{union2.1}, GOODS \cite{riess2007} and CANDELS/CLASH survey \cite{rodney2014,graur2014}. 

The distance modulus of SN-Ia can be derived from the observation of light curves through the empirical relation given by Tripp \cite{tripp}
\begin{equation}
\mu_{\mbox{\tiny SN}} = m^*_B + \alpha X_1 - \beta C - M_B + \Delta_M + \Delta_B, 	\label{tripp_formula}
\end{equation}

where $X_1$ and $C$ are the stretch and colour correction parameters, $m^*_B$ is the observed apparent magnitude and $M_B$ is the absolute magnitude in the 
B-band for a fiducial SN-Ia while $\alpha$ and $\beta$ are two nuisance parameters characterizing the luminosity-stretch, and luminosity-colour relations 
respectively. $\Delta_M$ is a distance correction based on the host-galaxy mass of the SN-Ia and $\Delta_B$ is a distance correction based on predicted biases 
from simulations. Usually, the nuisance parameters $\alpha$ and $\beta$ are simultaneously marginalized over with the cosmological parameters while assuming 
a particular background model. 

Adopting the BEAMS with Bias Corrections (BBC) \cite{beams} method, the nuisance parameters in the Tripp formula \eqref{tripp_formula} are retrieved and the 
observed distance modulus is reduced to the difference between the corrected apparent magnitude $m_B$ and the absolute magnitude $M_B$, as 

\begin{equation}
\mu_{\mbox{\tiny SN}} = m_B - M_B. \label{mu_SN}
\end{equation} 

Constraints on $M_B$ have been obtained by considering it a free parameter in our analysis. The distance modulus $\mu$ can be theoretically defined as

\begin{equation}
\mu = 5 \log_{10} d_L + 25,  \label{mu}
\end{equation}

where $d_L$ is the luminosity distance. With a bit of simple algebraic exercise, equation \eqref{mu} can be rewritten as

\begin{equation} \label{DL}
d_L(z)  = 10^{\frac{\mu - 25}{5}}.
\end{equation}

This $d_L$ is related to the comoving distance $d_C$ as

\begin{equation} \label{D_from_DL}
d_C(z) = \frac{d_L}{1+z}.	
\end{equation}

The total uncertainty matrix of distance modulus is given by 

\begin{equation} \label{pan_error}
\bm{\Sigma}_\mu = \mathbf{C}_{\mbox{\tiny stat}} + \mathbf{C}_{\mbox{\tiny sys}} ,
\end{equation}

where the statistical and systematic uncertainties, namely $\mathbf{C}_{\mbox{\tiny stat}}$ and $\mathbf{C}_{\mbox{\tiny sys}}$, are also included in our 
calculation.

\subsection{BAO data}

The Baryon Acoustic Oscillations are regular, periodic fluctuations in the matter power spectrum, and are widely used to measure distances in cosmology. For 
the purpose of our analysis, we have taken into account the volume-averaged BAO and the BAO-$H(z)$ data separately. The comoving sound horizon at photon drag 
epoch $r_d$ is considered as a free parameter in this work. The volume-averaged BAO data are utilized to constrain $r_d$, and the BAO $H(z)$ measurements are 
utilized for reconstructing $q$ in combination with the CC and SN datasets.

The volume-averaged BAO $\frac{D_V}{r_d}$ compilation consists of data from the Six-degree-Field Galaxy Survey (6dFGS) at $z=0.106$ \cite{beutler2011}, WiggleZ Dark 
Energy Survey at $z = 0.44$, $0.6$ and $0.73$ \cite{blake2012}, SDSS DR7 Main Galaxy Sample (MGS) at $z = 0.15$ \cite{ross2015}, LOWZ and CMASS samples of the DR12 
Baryon Oscillation Spectroscopic Survey (BOSS) galaxies at $z = 0.32, 0.57$ \cite{gilmarin2015} respectively, DR14 galaxy samples of the extended Baryon Oscillation 
Spectroscopic Survey (eBOSS) for Luminous Red Galaxies (LRGs) \cite{bautista2018} and quasars \cite{ata2018} samples at $z = 0.72, 1.52$, correlations of Ly$\alpha$ 
absorption in eBOSS DR14 galaxy sample at $z=2.34$ \cite{agathe} and cross-correlation of Ly$\alpha$ absorption and quasars in eBOSS DR14 galaxy sample at $z=2.35$ 
\cite{blomqvist}. For all the datasets mentioned, we appropriately use the covariance matrices that have been provided in the respective references.

The volume averaged distance is defined as

\begin{equation} \label{Dv}
D_V (z)= \left[d_C^2(z) \frac{c z}{H(z)}\right]^\frac{1}{3}.
\end{equation}

An alternative compilation of the Hubble data can be extracted from the radial BAO peaks in the galaxy power spectrum, or from the BAO peaks using the 
Ly-$\alpha$ forest of quasi stellar objects (QSOs), which are based on the clustering of galaxies or quasars. We utilize the latest compilation of the 
9 BAO $H(z)r_d$ measurements from different galaxy surveys, which includes the BOSS DR12 samples at 3 effective binned redshifts $z = 0.38, 0.51, 0.61$ 
\cite{alam2017}, eBOSS DR14 samples of LRGs and quasars at 4 effective redshifts $z = 0.98, 1.23, 1.52, 1.94$ \cite{zhao2019}, and the Ly$\alpha$ forest 
samples at $z=2.34$ \cite{agathe} and $z = 2.35$ \cite{blomqvist} respectively. We consider the $H(z)\frac{r_d}{r_{d, fid}}$ measurements along with the 
full covariance matrix, where the subscript `$fid$' stands for the fiducial value assumed in the process of acquiring these measurements in the respective 
data samples.

\subsection{RSD data}

Redshift-space distortions are an effect in observational cosmology where the spatial distribution of galaxies appears distorted due to the peculiar velocities 
of the galaxies causing a Doppler shift in addition to the redshift caused by the cosmological expansion. The growth of large structure can not only probe the 
background evolution of the universe, but also distinguish between different cosmological models which may have a similar background evolution but can stand in 
striking contrast to the growth of large scale structure in the universe. A recent compilation of the 63 RSD $f \sigma_8$ measurements, collected by \cite{rsd} 
and tabulated in \cite{li_rsd} is used for our analysis. This $f \sigma_8$ is called the growth rate of structure. The covariance matrix of the 63 $f\sigma_8$ 
data are assumed to be diagonal except for the WiggleZ DES, SDSS-III BOSS DR12 and SDSS-IV eBOSS DR14 galaxy sample subsets. The individual covariance matrices of 
the WriggleZ DES, SDSS-III BOSS DR12 and SDSS-IV eBOSS DR14 galaxy surveys are added to the $f{\sigma_8}$ error uncertainties for obtaining the full covariance 
matrix of the RSD dataset. 

As the $f\sigma_{8,\text{obs}}(z)$ data have been obtained assuming a fiducial $\Lambda$CDM cosmology \cite{rsd}, the Alcock-Paczynski (AP) effect \cite{APeffect} 
should be considered. A rough approximation of this AP effect \cite{macaulay2013, rsd, li_rsd} is given by

\begin{equation}
f\sigma_{8,\text{AP}}(z) \simeq \frac{H(z) d_A(z)}{H^{\text{fid}}(z,\Omega_{m,0}) d_A^{\text{~fid}}(z,\Omega_{m,0}) } f\sigma_{8,\text{obs}}(z), 
\label{APfactor}
\end{equation}

where $d_A(z) = \frac{d_C}{(1+z)}$ is the angular diameter distance. This AP correction has been found to have little effect on the mean values of $f \sigma_{8}(z)$ 
\cite{li_rsd}. Thus, we proceed with our analysis directly utilizing the observed $f\sigma_{8}$ dataset, similar to \cite{zhang_li, yang2021, benisty}.

\subsection{$H_0$ measurement} \label{H0-priors}

Different strategies for determining value of $H_0$ is well known in the recent literature \cite{cosmoletter2}. Locally, the Hubble parameter has been measured 
to be $H_0 = 73.2 \pm 1.3$ km s$^{-1}$ Mpc$^{-1}$ obtained from the expanded sample of 75 Milky Way Cepheids with Hubble Space Telescope (HST) photometry and 
Gaia EDR3 parallaxes by the SH0ES team \cite{riess2021} (hereafter referred to as R21).

Another strategy involves an extrapolation of data on the early Universe from the Cosmic Microwave Background (CMB) which yields, $H_0 = 67.27 \pm 0.60$ km s$^{-1}$ 
Mpc$^{-1}$ provided by Planck 2020 power spectra (TT, TE, EE+lowE) measurements \cite{planck2020}, assuming a base $\Lambda$CDM model (referred to as P20).

In what follows, the reconstruction is also undertaken with these values of $H_0$ being included to the CC Hubble parameter compilation. This exercise has been 
done in order to check if there is any qualitative change in the results due to the introduction of these priors.

\subsection{$\Omega_{k,0}$ measurement}

The $\Lambda$CDM model assumes that the spatial hyper-surfaces are flat. This is a prediction that can be tested to high accuracy by a combination of the Planck 
likelihood with the CMB power spectra. A combination of the Planck 2020 temperature and polarization power spectra with lensing data gives $\Omega_{k,0} = -0.0106 
\pm 0.0065$ (TT, TE, EE+lowE+lensing) \cite{planck2020}. We shall investigate the effect of this non-zero spatial curvature prior on the reconstruction of $q$.

\section{Gaussian Process methodology} \label{sec-gp}

Gaussian Process (GP) is a non-parametric method for reconstructing a function without considering any \textit{apriori} parametrization ansatz. A GP is a 
distribution over functions, generalizing the idea of a Gaussian probability distribution for a finite collection of datasets. Assuming a Gaussian distributed 
observational data set, we employ the GP method to obtain the most probable function describing this data along with the associated error uncertainties. GPs 
also provide a robust way to estimate derivatives of this aforesaid function. As an example, we consider a compilation of the comoving distances $D$ at different 
redshifts obtained from SN-Ia observations. GPs are capable of directly reconstructing the underlying continuous function $D(z)$ along with its derivatives. 
The individual posterior distributions of the reconstructed functions $D(z)$ and its higher derivatives can be expressed as a joint Gaussian distribution of 
different data sets of the comoving distance $D$. Incorporating the reduced Hubble parameter measurements from the CC and BAO Hubble datasets provides additional 
constraints on the first-order derivative of $D(z)$ in our analysis, which reduces uncertainties in the reconstructed functions $D(z)$ and $D'(z)$ at high $z$.

GPs are characterised by a mean function $\mu(z)$ and a covariance function $\kappa(z, \tilde{z})$. The latter depends on a set of {\it hyperparameters}, namely 
the characteristic length scale $l$ and the signal variance $\sigma_f$, which correlates the values of the reconstructed function at the redshift $z$ and some 
other redshift $\tilde{z}$, in the neighbourhood of $z$. The standard choices for $\kappa (z, \tilde{z})$ are the squared exponential and the Mat\'{e}rn $\nu$ 
class covariance function (see \cite{rw} for a comprehensive discussion). The squared exponential covariance is defined as 

\begin{equation}
\kappa(z, \tilde{z}) = \sigma_f^2 \exp \left[ - \frac{(z-\tilde{z})^2}{2l^2}\right] ,
\end{equation}

and is extensively used in cosmology. For a reconstruction involving an $n$th order derivative, the Mat\'{e}rn $\nu$ covariance works well if $\nu > n$ 
\cite{seikel2013}. We refer to Ref. \cite{holsclaw2,gp_seikel,gp_ref26,gp_ref27,gp_ref28,purba_cddr,gp_ref29,purba_int,purba_j,gp_ref30,ryan,yang2021,
wang-meng,wang-meng2,zhou-peng,cai-saridakis,manos,benisty,keeley2020,arjona,xia,bilicki,lin,jesus_nonpara,levisaid2021,zhang_li,seikel2013,haridasu} 
for elaborate discussions on the various applications of GP in cosmology. For a general overview, one can refer to the GP 
website\footnote{\url{http://www.gaussianprocess.org}}.

Throughout this work we consider a zero mean $\mu(z)=0$ and the Mat\'{e}rn $9/2$ covariance function. The Mat\'{e}rn $9/2$ covariance is given by 

\begin{equation}
\begin{split}
\kappa(z,\tilde{z}) = \sigma_f^2 e^{\left( \frac{-3 \vert z - \tilde{z} \vert}{l} \right)} \left[ 1 + \frac{3 \vert z - \tilde{z} \vert}{l} +  \right. ~~~~~~~~~~~~\\
\left. + \frac{27 \left( z - \tilde{z} \right)^2}{7l^2} + \frac{18 \vert z - \tilde{z} \vert ^3}{7l^3} + \frac{27 \left( z - \tilde{z} \right)^4}{35l^4}\right]. \label{mat92}
\end{split}
\end{equation} 

According to \cite{haridasu} a constant value for the mean $\mu(z)$ does not play any significant role in obtaining the reconstructions, as it only remains 
an additive factor for the final predictions. Moreover, assuming an explicit functional form for $\mu(z)$ compromises the model-independent nature of the 
reconstruction as a prior functional form is being imposed. As a result the reconstructed posterior mean tends to the prior mean. Therefore, a zero mean prior 
is comparatively a safe choice in comparison to other asymmetric priors to make predictions, unless an accurate prior knowledge of the target function is 
available. For an extensive study on the effect of a non-zero $\mu(z)$ we refer to \cite{haridasu,eoin}. 

For reconstructing any function via GP, the hyperparameters $\sigma_f$ and $l$ needs to be estimated. They can be trained by maximizing the marginal likelihood, 
which is a marginalization over function values $\left\lbrace D(z_i) \right\rbrace$ at redshift locations $\left\lbrace zi \right\rbrace_{i=1}^{N}$. The log-marginal 
likelihood on assuming a Gaussian prior with zero mean is given by

\begin{equation} \label{likelihood}
\begin{split}
\ln \mathcal{L} = -\frac{1}{2} \mathbf{D}^{\mbox{\small T}} \left[ K(\mathbf{Z}, \mathbf{Z}) + \bm{\mathcal{C}} \right]^{-1} \mathbf{D} + ~~~~~~~~\\ 
-\frac{1}{2} \ln\vert K(\mathbf{Z}, \mathbf{Z})+\bm{\mathcal{C}} \vert -\frac{N}{2}\ln 2\pi,
\end{split}
\end{equation} 

where $K(\mathbf{Z},\mathbf{Z})$ is the covariance matrix given by $[K(\mathbf{Z},\mathbf{Z})]_{ij} = \kappa(z_i, z_j)$ at $\mathbf{Z}= \{z_i\}_{i=1}^N$ 
observational redshift points and $\bm{\mathcal{C}}$ is the covariance matrix of the data.

Besides estimating the GP hyperparameters, one also needs to constrain the nuisance parameters involved in the datasets for a self-consistent reconstruction. 
The publicly available \texttt{GaPP}\footnote{\url{https://github.com/carlosandrepaes/GaPP}} (Gaussian Processes in Python) code developed by Seikel et al. 
\cite{gp_seikel} has been modified and utilized accordingly in our work.

\section{Reconstruction from Background data}

The deceleration parameter $q$ defined in Eq. \eqref{qdef} can be written as a function of redshift $z$ as

\begin{equation}
q(z) = -1 + (1+z) \frac{H'}{H} = -1 + (1+z) \frac{E'}{E}  , \label{qz}
\end{equation} 

where a `prime' denotes derivative with respect to the redshift $z$.

One can obtain a relation between the reduced Hubble parameter $E(z)$ and the normalised comoving distance $D(z)$ from equations \eqref{D} and 
\eqref{D_from_dc}, such that

\begin{equation} \label{D'}
E(z)=\frac{\sqrt{1+\Omega_{k,0}D^2}}{D'(z)}.
\end{equation} 

Finally, $q$ can be represented as a function of the normalised comoving distance $D$ and its derivatives as

\begin{equation} 
q(z) = -1 + \frac{\Omega_{k,0}DD'^2 - (1+\Omega_{k,0}D^2)D''}{D'(1+\Omega_{k,0}D^2)}(1+z) .	 \label{qnew}
\end{equation}

This will serve as the key equation for the non-parametric reconstruction of $q(z)$ using different combinations of the background datasets.

\begin{figure}
	\begin{center}
		\includegraphics[angle=0, width=0.5\textwidth]{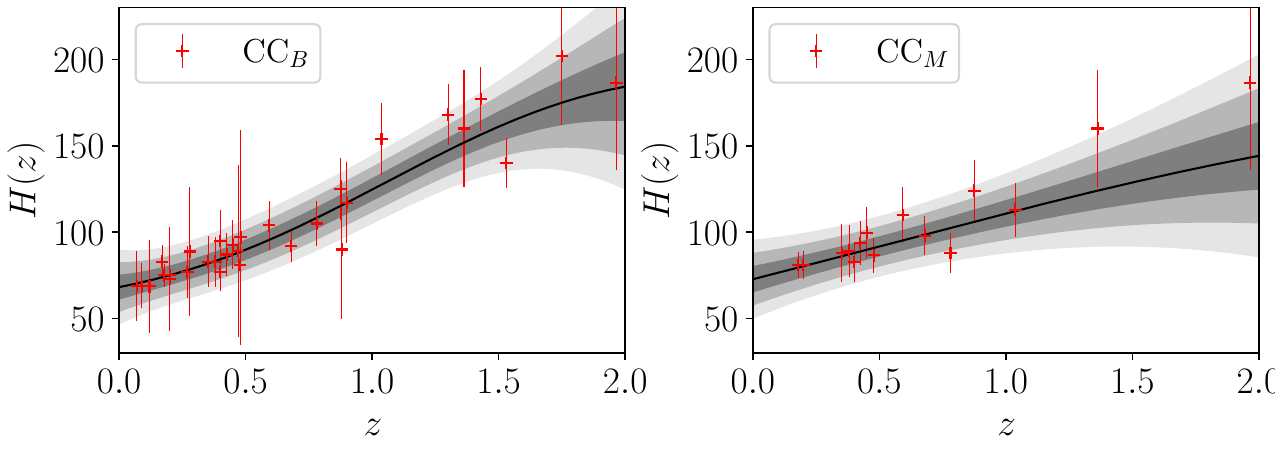}
	\end{center}
	\caption{{\small Plots for GP reconstructed $H(z)$ from the updated datasets CCB (left) and CCM (right) respectively. The solid black line represents 
			the mean values of the reconstructed $H(z)$. }} \label{H_cc_plot}
\end{figure}

\begin{table}
	\caption{{\small Table showing the GP reconstructed $H_0$ (in units of km Mpc$^{-1}$ s$^{-1}$) from the latest updated CC data compilation.}}
	\begin{center}
		\resizebox{0.5\textwidth}{!}{\renewcommand{\arraystretch}{1.5} \setlength{\tabcolsep}{20pt}\centering 
			\begin{tabular}{l  c  c} 
				\hline \hline 
				&   CCB  & CCM  \\
				\hline
				$H_0$ ~~~~&  $68.193 \pm 7.209$ ~~~~~& $72.776 \pm 7.636$ \\
				\hline \hline
			\end{tabular}
		}
	\end{center}
	\label{H0_cc_result}
\end{table}

The reconstruction of $q$, in the present work, involves a two-step analysis. In the first step, we obtain the marginalized constraints on $M_B$ and $r_d$. 
In the second step, these constraints are utilized in reconstructing $D(z)$, $D'(z)$, and $D''(z)$ for the corresponding dataset combination. Finally, the 
deceleration parameter $q(z)$ is derived by using the reconstructed comoving distance $D(z)$, its derivatives $D'(z)$, and $D''(z)$ according to equation 
\eqref{qnew}.

\subsection{Constraints on $M_B$ and $r_d$} \label{sec5.1}

\begin{table*}
	\caption{{\small Table showing the marginalised constraints on $M_B$ and $r_d$ (in units of Mpc) for different combinations of datasets.}}
	\begin{center}
		\resizebox{\textwidth}{!}{\renewcommand{\arraystretch}{1.5} \setlength{\tabcolsep}{18pt}\centering 
			\begin{tabular}{l  c  c  c  c  c } 
				\hline
				\hline 
				& & CCB+SN & CCB+SN+BAO & CCM+SN & CCM+SN+BAO \\
				\hline 
				& $M_B$  & $-19.409 \pm 0.010$  & $-19.412 \pm 0.007$ &  $-19.341 \pm 0.011$  & $-19.390 \pm 0.008$ \\[-0.75em]
$\Omega_{k,0}=0$ & \\[-0.75em]
				& $r_d$  & -	  &  $148.76 \pm 0.28$  &  	-   &   $149.61 \pm 0.39$  \\
				\hline 
				& $M_B$ &  $-19.412 \pm 0.014$  & $-19.413 \pm 0.009$ &  $-19.353 \pm 0.015$  & $-19.412 \pm 0.010$ \\[-0.75em]
$\Omega_{k,0}\neq 0$ & \\[-0.75em]
				& $r_d$ &  -	  &  $148.67 \pm 0.33$  &  	-   &   $150.22 \pm 0.47$  \\
				\hline
				\hline
			\end{tabular}
		}
	\end{center}
	\label{MB_rd_tab}
\end{table*}

We begin with a GP reconstruction of the Hubble parameter from the latest CC measurements. The systematic errors due to the initial mass function (IMF) 
and stellar population synthesis (SPS) models, namely the BC03 and MaStro models, associated with the CC data were recently analysed in \cite{sps2}. The 
systematic errors linked with the CC data have been added to the covariance matrices of the current CC data similar to the procedure in Ref. \cite{lin_cc}. 
Columns (2) and (5) in Table 3 of \cite{sps2} accounts for these two systematic errors. We interpolate these two columns to get the error budget of the 
current measurements at each redshift due to these two extra sources. The covariance matrices, $\text{Cov}^{\text{IMF}}_{i,j}$ and $\text{Cov}_{i,j}^{\text{SPS}}$ 
are obtained according to Eq. (9) in \cite{sps2}, as
\begin{equation}\label{eq:Cov-CC-stimated}
{\text{Cov}}^{\text{X}}_{i,j}=\widehat{\eta^{\text{X}}}(z_i)H(z_i)\widehat{\eta^{\text{X}}}(z_j)H(z_j) ,
\end{equation}

where $\widehat{\eta^{\text{X}}}(z)$'s are obtained by interpolation with the data provided in Table 3 of \cite{sps2}, and $H(z_i)$'s are CC measurements 
at different redshifts. The covariance matrices, $\text{Cov}^{\text{IMF}}_{i,j}$ and $\text{Cov}_{i,j}^{\text{SPS}}$, added to the statistical uncertainties for 
obtaining the total covariance matrix of the current CC dataset. Plots for the reconstructed $H(z)$ from the updated CCB and CCM samples are shown in Fig. 
\ref{H_cc_plot}. The reconstructed $H_0$ values obtained from the individual CCB and CCM samples, shown in Table \ref{H0_cc_result}, are utilized solely 
for obtaining the constraints on parameters $M_B$ and $r_d$.

With the smooth reconstructed function $H(z)$ from the CC data, we use a composite trapezoidal rule \cite{trapez} to obtain the integral 

\begin{eqnarray} 
	\mathcal{I} &=& \int_{0}^{z} \frac{\dif z'}{H(z')} , \nonumber \\ 
	&\simeq& \frac{1}{2} \sum_{i=0}^{n-1} (z_{i+1} - z_i) \left[\frac{1}{H(z_{i+1})}+\frac{1}{H(z_{i})}\right]. \label{integral_H}
\end{eqnarray}

The numerical error associated with $\mathcal{I}$ is of order $10^{-6}$, and does not adversely affect our analysis. The statistical uncertainty in 
$\mathcal{I}$ is obtained by the error propagation formula

\begin{equation}  \label{integral_sigH}
\sigma^2_{\mathcal{I}} = \sum_{i=0}^{n} \frac{1}{4} (z_{i+1} - z_i)^2 \left[\frac{\sigma^2_{H_{i+1}}}{H^4_{i+1}}+\frac{\sigma^2_{H_{i}}}{H^4_{i}}\right]. 
\end{equation}

Using equations \eqref{integral_H} and \eqref{integral_sigH}, we obtained a smooth function of the comoving distance $d_C$ and its associated uncertainty 
$\sigma_{d_C}$ from the CC Hubble data as

\begin{eqnarray} \label{D_recon_cc}
{d_C}_{\mbox{\tiny CC}}= \begin{cases}
\frac{c}{H_0 \sqrt{\Omega_{k,0}}}\sinh \left[  H_0 \sqrt{\Omega_{k,0}} \mathcal{I}(z)\right] & \Omega_{k,0}>0, \\
c ~\mathcal{I}(z) & \Omega_{k,0}=0, \\
\frac{c}{ H_0 \sqrt{-\Omega_{k,0}}}\sin \left[  H_0 \sqrt{-\Omega_{k,0}} \mathcal{I}(z)\right] &  \Omega_{k,0}<0. 
\end{cases}
\end{eqnarray} 

The error $\sigma_{d_C}$ associated with the reconstructed $d_C$ from the CC Hubble data is

\begin{eqnarray} \label{sigD_recon_cc}
{\sigma_{d_C}}_{\mbox{\tiny CC}}= \begin{cases}
c \cosh \left[ H_0 \sqrt{\Omega_{k,0}} ~ \mathcal{I}(z)\right] \sigma_{\mathcal{I}}(z) & \Omega_{k,0}>0, \\
c ~\sigma_{\mathcal{I}}(z) & \Omega_{k,0}=0, \\
c \cos \left[ H_0 \sqrt{-\Omega_{k,0}} ~ \mathcal{I}(z)\right] \sigma_{\mathcal{I}}(z) &  \Omega_{k,0}<0.
\end{cases}
\end{eqnarray}

The reconstructed ${d_C}_{\mbox{\tiny CC}}$ takes the role of a theoretical model which are further utilized to obtain the distance modulus from the 
CC Hubble data $\mu_{\mbox{\tiny CC}}$ using Eq. \eqref{mu} as 

\begin{equation}
	\mu_{\mbox{\tiny CC}} = 5 \log_{10} \left[{d_C}_{\mbox{\tiny CC}}(1+z)\right] + 25 . \label{mu_CC}
\end{equation}

The associated 1$\sigma$ uncertainty $\sigma _{\mu_{\mbox{\tiny CC}}}$ is given by 

\begin{equation} 
	\sigma _{\mu_{\mbox{\tiny CC}} }=\frac{5}{\ln 10}\frac{{\sigma _{d_C}}_{\mbox{\tiny CC}}}{{d_C}_{\mbox{\tiny CC}}} . \label{sigma_mu_CC}
\end{equation}

The distance modulus from the Pantheon SN compilation are combined with the CC $H(z)$ measurements to account for the degeneracy between the absolute magnitude 
$M_B$ of SN-Ia and the Hubble parameter at present epoch $H_0$. Instead of setting a fiducial value $M_B = -19.35$ corresponding to the reference $\Lambda$CDM 
model as done in \cite{lin}, we reconstruct the corrected apparent magnitudes $m_B$ adopting a GP regression and obtain the constraints on $M_B$ by minimizing 
the $\chi^2$ function 

\begin{equation}
\chi^2 = \Delta \bm{\mu}^{\mbox{\small T}} \cdot \bm{\Sigma}^{-1} \cdot \Delta \bm{\mu}. \label{chi2}
\end{equation}

Here $\Delta \bm{\mu} = \bm{\mu}_{\mbox{\tiny SN}}- \bm{\mu}_{\mbox{\tiny CC}}$ and $\bm{\Sigma} = \bm{\Sigma}_{\mu_{\mbox{\tiny SN}}} + \sigma^2_{\mu_{\mbox{\tiny 
CC}}}$ respectively. We get the best fit constraints on $M_B$ and the associated 1$\sigma$ uncertainties by a Markov Chain Monte Carlo (MCMC) analysis with the 
assumption of a uniform prior distribution for $M_B \in [-25, -15]$ so that any initial dependence on the $\Lambda$CDM model is eliminated.

In order to introduce the BAO $H r_d$ measurements in combination with the CC and Pantheon data, we need to obtain the constraints on $r_d$ independent of any fiducial 
background cosmological model. The volume-averaged BAO data are utilized for this purpose. We reconstruct $\frac{D_V}{r_d}$ via another GP and obtain the joint constraints 
on $M_B$ and $r_d$. One can evaluate the comoving distances from the reconstructed volume-averaged BAOs in combination with the reconstructed CC Hubble data, by means of 
Eq. \eqref{Dv}.

\begin{equation} \label{D_recon_bao}
{d_C}_{\mbox{\tiny BAO}}  = \left[\frac{D_V^3(z) H(z)}{c z}\right]^\frac{1}{2} .
\end{equation}

\begin{table*}[h!]
	\caption{{\small Table showing the inferred values of $H_0$ (in units of km Mpc$^{-1}$ s$^{-1}$) for different combinations of datasets computed from equation 
			\eqref{H0_gp} .}}
	\begin{center}
		\resizebox{\textwidth}{!}{\renewcommand{\arraystretch}{1.5} \setlength{\tabcolsep}{20pt}\centering 
			\begin{tabular}{c  c  c  c  c  c} 
				\hline
				\hline 
				&  & CCB+SN & CCB+SN+BAO & CCM+SN & CCM+SN+BAO \\
				\hline
				$\Omega_{k,0} = 0$  & $H_0$ &  $ 68.711 \pm 0.414 $  & $ 68.395 \pm 0.412 $ &  $ 70.636 \pm 0.425 $  & $ 69.028 \pm 0.413 $ \\
				\hline
				$\Omega_{k,0} \neq 0$ & $H_0$ & $ 68.397 \pm 0.415 $  & $ 68.396 \pm 0.413 $ &  $ 70.638 \pm \pm 0.428 $  & $ 68.395 \pm 0.416 $ \\
				\hline
				\hline
			\end{tabular}
		}
	\end{center}
	\label{H0_gp_result}
\end{table*}

This reconstructed ${d_C}_{\mbox{\tiny BAO}}$ along with its 1$\sigma$ uncertainty are further treated following a similar manner as in Eq. \eqref{mu_CC}, 
\eqref{sigma_mu_CC} and \eqref{chi2} to simultaneously constrain $M_B$ and $r_d$ via a minimization of the combined $\chi^2$, employing another MCMC analysis 
assuming a uniform prior distribution with $r_d \in [135, 160]$. We adopted a python implementation of the ensemble sampler for MCMC, the publicly available \texttt{emcee}\footnote{\url{https://github.com/dfm/emcee}}, introduced by Foreman-Mackey et al. \cite{emcee}. The best-fit results of $M_B$ and $r_d$ along 
with their respective 1$\sigma$ uncertainties are given in Table \ref{MB_rd_tab}. To incorporate the influence of spatial curvature, we first consider it to 
be zero and later assign $\Omega_{k,0} = -0.0106 \pm 0.0065$ from the Planck 2020 (TT,TE,EE+lowE+lensing) \cite{planck2020} probe.

\subsection{Reconstructing $D(z)$ and its derivatives} \label{sec-5.2}

The marginalized $M_B$ constraints from Table \ref{MB_rd_tab} are substituted in Eq. \eqref{D_from_DL} for computing the comoving distances $d_C$ of all 
supernovae in the Pantheon compilation via a transformation from $\mu$ using Eq. \eqref{DL}. The uncertainty matrix $\bm{\Sigma}_{d_C}$ associated with 
the SN-Ia comoving distance data is obtained from the total uncertainty matrix of distance modulus $\bm{\Sigma}_\mu$ given in Eq. \eqref{pan_error}. We 
identify the comoving distances $d_C$ as the training dataset that spans the function space. We then exercise a GP regression of the SN comoving distances, 
and reconstruct the target functions $d_C(z)$ and ${d_C}^\prime(z)$ incorporating the theoretical condition $d_C(z = 0) = 0$ with uncertainty zero. Being 
directly measured from SN-Ia, these $d_C$ are independent of $H_0$. We infer $H_0$ utilizing Eq. \eqref{D'} as 

\begin{equation} \label{H0_gp}
H_0 = c{{\left[{{d_C}^\prime}^2(0) - \Omega_{k,0} {d_C}^2(0)\right]}}^{-\frac{1}{2}}.
\end{equation} 

The uncertainty associated with the inferred $H_0$ are propagated from the uncertainties associated with $d_C(0)$ and ${d_C}^\prime(0)$ and $\Omega_{k,0}$ 
respectively. The inferred $H_0$ values obtained from Eq. \eqref{H0_gp} are shown in Table \ref{H0_gp_result}.

For computing the Hubble parameter from the BAO $H r_d$ measurements, we substitute the marginalized $r_d$ constraints from Table \ref{MB_rd_tab}. The resulting 
values of the Hubble parameter obtained are added with the CC $H(z)$ measurements to form the CC+BAO Hubble data. The total covariance matrix is obtained by 
appending the individual CC and BAO covariance matrices corresponding to the full $H(z)$ sample.

After preparation of CC and CC+BAO Hubble datasets, we normalize them with inferred values of $H_0$ from Table \eqref{H0_gp_result} to obtain the reduced 
Hubble parameter $E$. Considering the error associated with the inferred $H_0$ to be $\sigma_{H_0}$, one can calculate the uncertainty covariance matrix 
associated with $E$, i.e. $\bm{\Sigma}_{E}$ as,  

\begin{equation} \label{sigE_recon}
\bm{\Sigma}_{E} = \frac{\bm{\Sigma}_H}{ {H_0}^2} + \frac{H^2}{{H_0}^4}{\sigma_{H_0}}^2 , 
\end{equation} 

where $\bm{\Sigma}_{H}$ is the uncertainty covariance matrix of the Hubble data compilation. The comoving distances from the Pantheon compilation are normalised 
with the same inferred $H_0$ values given in Table \eqref{H0_gp_result} to obtain the dimensionless comoving distances $D$ using Eq. \eqref{D_from_dc}. The 
uncertainty associated with training dataset $D$ are propagated from the uncertainties of $\mu$ ($\bm{\Sigma}_\mu$ in Eq. \eqref{pan_error}) and $H_0$ $(\sigma_{H_0})$ 
via the standard error propagation formula. Eventually, the normalised comoving distances are combined with the reduced Hubble parameter measurements via equation 
\eqref{D'} as additional constraints on the first-order derivative of $D(z)$, i.e. $D'(z)$, in our analysis.

\begin{figure*}[t!]
	\begin{center}
		\includegraphics[angle=0, width=0.8\textwidth]{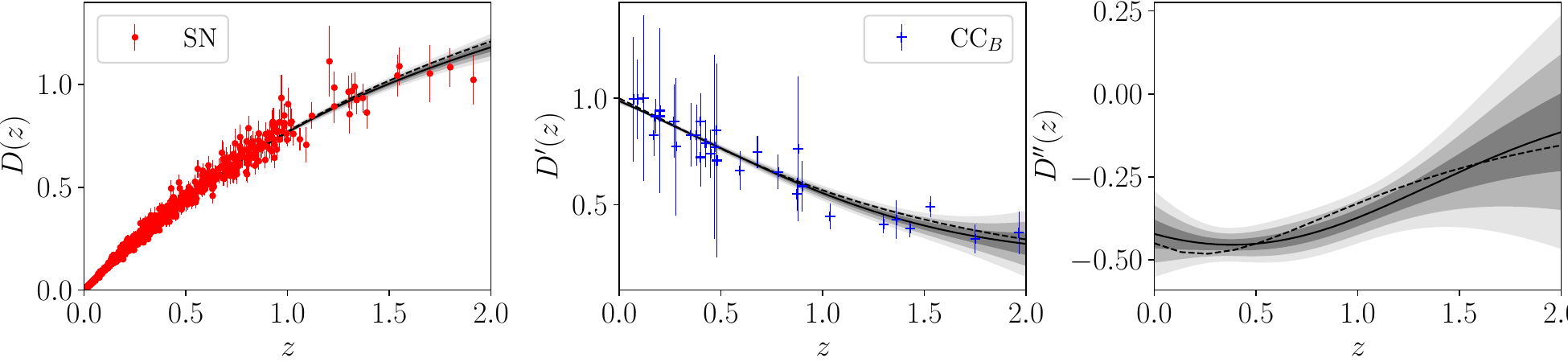}
	\end{center}
	\caption{{\small Plots for the reconstructed dimensionless comoving distance $D(z)$, its derivatives $D'(z)$ and $D''(z)$ using combined CCB+SN data (Set N1) for  
			a spatially flat universe ($\Omega_{k,0}=0$). The black solid line is the mean curve. The associated 1$\sigma$, 2$\sigma$ and 3$\sigma$ confidence regions 	
			are shown in lighter shades. The specific markers with error bars represent the observational data. The $\Lambda$CDM model with $\Omega_{m,0} = 0.3$ is 
			represented by the dashed line.}} \label{D_N1}
\end{figure*}

\begin{figure*}[t!]
	\begin{center}
		\includegraphics[angle=0, width=0.8\textwidth]{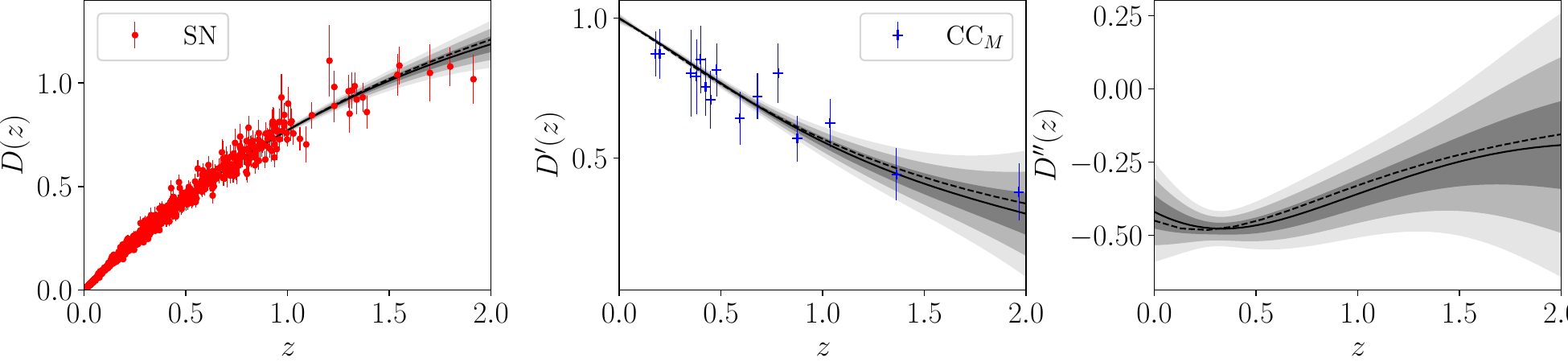}
	\end{center}
	\caption{{\small Plots for the reconstructed dimensionless comoving distance $D(z)$, its derivatives $D'(z)$ and $D''(z)$ using combined CCM+SN data (Set N2) for 
			a spatially flat universe ($\Omega_{k,0}=0$). The black solid line is the mean curve. The associated 1$\sigma$, 2$\sigma$ and 3$\sigma$ confidence regions 
			are shown in lighter shades. The specific markers with error bars represent the observational data. The $\Lambda$CDM model with $\Omega_{m,0} = 0.3$ is 
			represented by the dashed line.}} \label{D_N2}
\end{figure*}

\begin{figure*}[t!]
	\begin{center}
		\includegraphics[angle=0, width=0.8\textwidth]{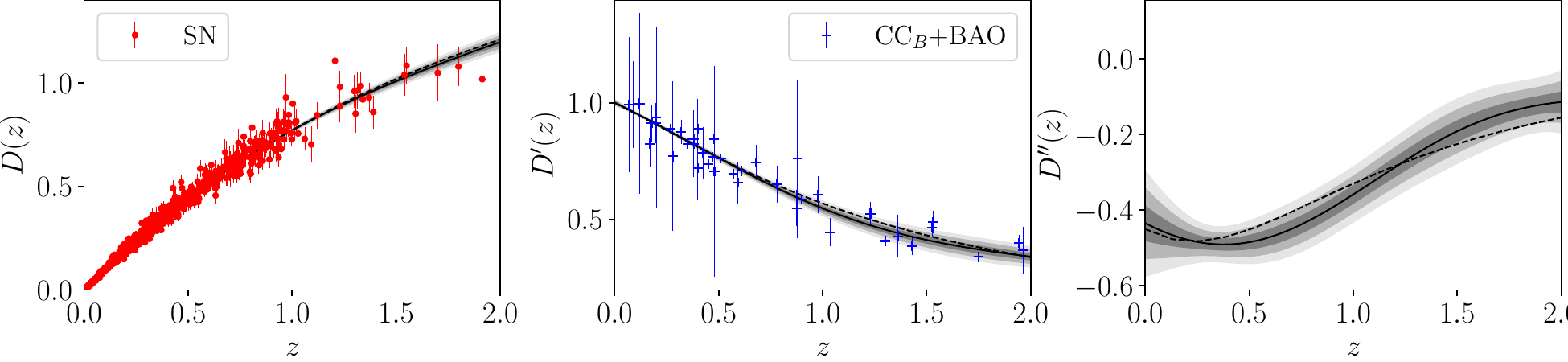}
	\end{center}
	\caption{{\small Plots for the reconstructed dimensionless comoving distance $D(z)$, its derivatives $D'(z)$ and $D''(z)$ using combined CCB+SN+BAO data (Set N3) for 
			a spatially flat universe ($\Omega_{k,0}=0$). The black solid line is the mean curve. The associated 1$\sigma$, 2$\sigma$ and 3$\sigma$ confidence regions are 
			shown in lighter shades. The specific markers with error bars represent the observational data. The $\Lambda$CDM model with $\Omega_{m,0} = 0.3$ is represented 
			by the dashed line.}} \label{D_N3}
\end{figure*}

\begin{figure*}[t!]
	\begin{center}
		\includegraphics[angle=0, width=0.8\textwidth]{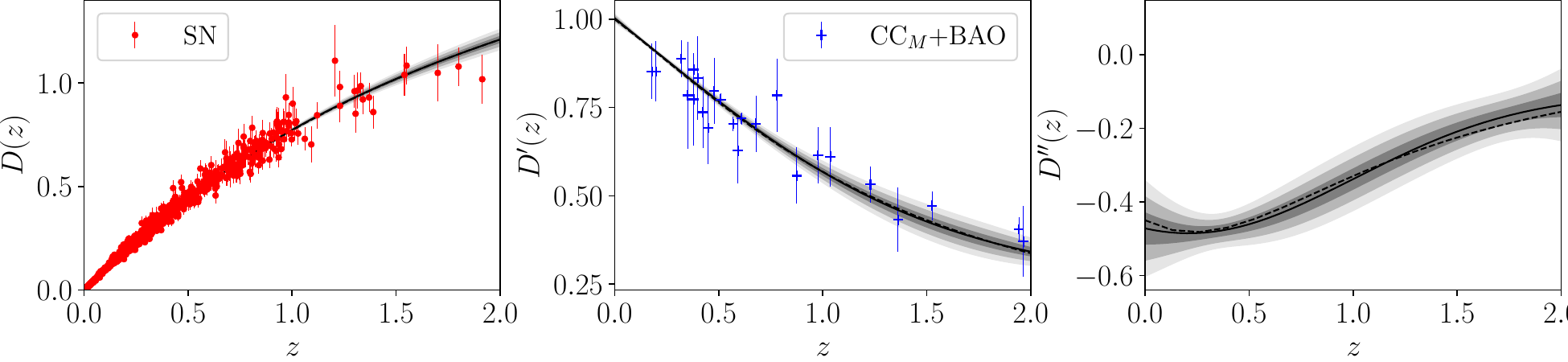}
	\end{center}
	\caption{{\small Plots for the reconstructed dimensionless comoving distance $D(z)$, its derivatives $D'(z)$ and $D''(z)$ using combined CCM+SN+BAO data (Set N4) for 
			a spatially flat universe ($\Omega_{k,0}=0$). The black solid line is the mean curve. The associated 1$\sigma$, 2$\sigma$ and 3$\sigma$ confidence regions are 
			shown in lighter shades. The specific markers with error bars represent the observational data. The $\Lambda$CDM model with $\Omega_{m,0} = 0.3$ is represented 
			by the dashed line.}} \label{D_N4}
\end{figure*}

\begin{figure*}[h!]
	\begin{center}
		\includegraphics[angle=0, width=\textwidth]{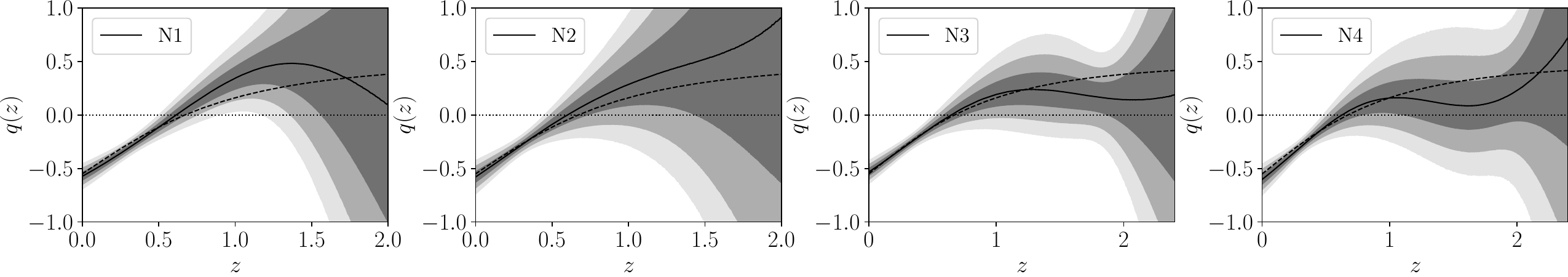}
	\end{center}
	\caption{{\small Plots for $q(z)$ reconstructed from the combined datasets N1, N2, N3, N4 for a spatially flat universe ($\Omega_{k,0}=0$). The solid black line 
			represents the mean values of the reconstructed $q(z)$. The black dashed line shows $q(z)$ corresponding to the $\Lambda$CDM model with $\Omega_{m,0} = 
			0.3$.}} \label{qplot}
\end{figure*}

\begin{figure*}[h!]
	\begin{center}
		\includegraphics[angle=0, width=\textwidth]{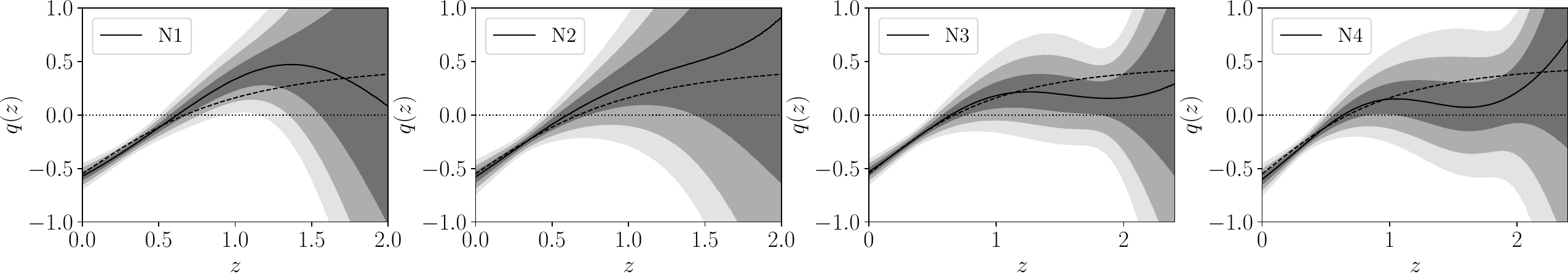}
	\end{center}
	\caption{{\small Plots for $q(z)$ reconstructed from the combined datasets N1, N2, N3, N4 for a universe with a non-zero spatial curvature given by the Planck 
			2020 measurement $\Omega_{k,0}= -0.0106 \pm 0.0065$ \cite{planck2020}. The solid black line represents the mean values of the reconstructed $q(z)$. The 
			black dashed line shows $q(z)$ corresponding to the $\Lambda$CDM model with $\Omega_{m,0} = 0.3$.}} \label{qplot_ok}
\end{figure*}

\begin{table*}[h!]
	\caption{{\small Table showing the reconstructed mean values along with the 1$\sigma$ uncertainties for $q_0$ corresponding to the datasets N1, N2, N3 and N4. 
			An estimate for the late-time deceleration-acceleration transition redshift $z_t$ is also provided.}}
	\begin{center}
		\resizebox{\textwidth}{!}{\renewcommand{\arraystretch}{1.8} \setlength{\tabcolsep}{24 pt} \centering  
			\begin{tabular}{l c c c c c c }
				\hline \hline
				&  &  N1 & N2 & N3 & N4  \\ 
				\hline 
				& $q_0$ & $-0.573^{ +0.041}_{-0.042}$ & $-0.580^{+0.055}_{-0.063}$ &  $-0.533^{+0.038}_{-0.038}$  & $-0.574^{+0.044}_{-0.045}$ \\ [-0.95em]
$\Omega_{k,0} = 0$ & \\[-0.95em]
				& $z_t$ & $0.611_{-0.045}^{+0.065}$ & $0.601_{-0.071}^{+0.140}$ &  $0.644_{ -0.064}^{ +0.092}$  & $0.602_{-0.050}^{+0.065}$ \\ 
				\hline 
				& $q_0$ &  $-0.571^{ +0.043 }_{ -0.044}$ & $-0.573^{+0.062}_{ -0.062}$ &  $-0.532^{ +0.041}_{-0.041}$  & $-0.573^{ +0.047}_{ -0.048}$ \\ [-0.95em]
$\Omega_{k,0} \neq 0$ & \\[-0.95em]
				& $z_t$ &  $0.621_{-0.046}^{+0.066}$ & $0.605_{-0.081}^{+0.182}$ &  $0.643_{-0.069}^{+0.094}$  & $0.610_{-0.055}^{+0.070}$  \\ 
				\hline 
				\hline 
			\end{tabular} 
		}
	\end{center}
	\label{q0_tab}
\end{table*}

Thus, having acquired all the necessary training data (in this case the SN comoving distance $D$ and reduced Hubble parameter $E$) for the GP analysis we 
proceed with a non-parametric reconstruction of the normalised comoving distance $D(z)$ and its derivatives $D'(z)$ and $D''(z)$ at different redshift $z$, 
as described in Sec. \ref{sec-gp} for the following combination of datasets. 

\begin{itemize}
	\item Set N1 - CCB+SN, \item Set N2 - CCM+SN, \item Set N3 - CCB+SN+BAO, \item Set N4 - CCM+SN+BAO.
\end{itemize}

The hyperparameters in the Mat\'{e}rn 9/2 covariance function, defined in \eqref{mat92} are obtained by marginalizing the log-likelihood function (see Eq. 
\eqref{likelihood}). Utilizing the trained hyperparameters, we reconstruct the mean values for the most probable continuous function $D(z)$ of the distance 
data and its derivatives, along with the associated confidence levels. Plots for the reconstructed $D(z)$, $D'(z)$ and $D''(z)$ versus $z$ are shown in Fig. 
\ref{D_N1}, \ref{D_N2}, \ref{D_N3} and \ref{D_N4} for N1, N2, N3 and N4 dataset combinations, respectively.

\begin{figure*}[h!]
	\begin{center}
				\includegraphics[angle=0, width=\textwidth]{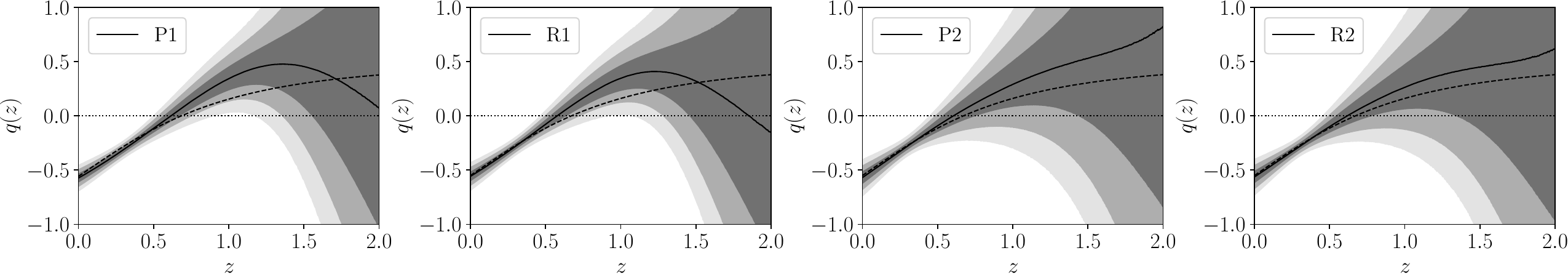}\\
				\includegraphics[angle=0, width=\textwidth]{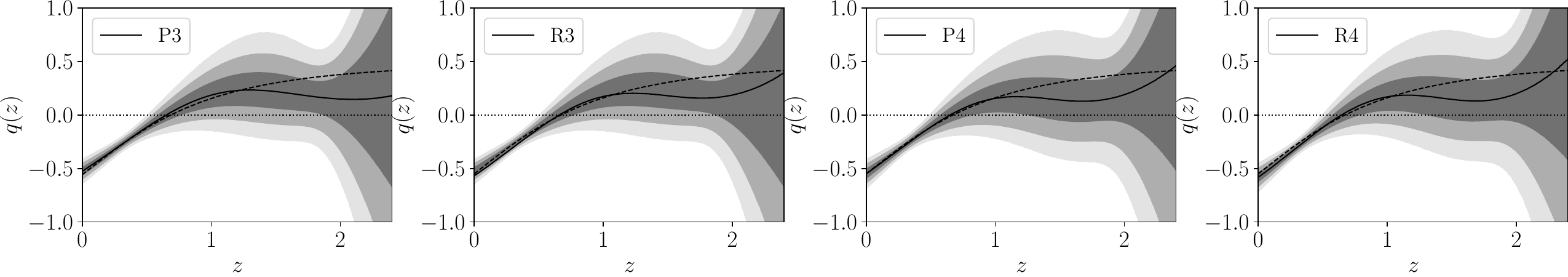}
	\end{center}
	\caption{{\small Plots for $q(z)$ reconstructed from the combined datasets P1, R1, P2, R2, P3, R3, P4 and R4 for a spatially flat universe ($\Omega_{k,0}
			=0$). The solid black line represents the mean values of the reconstructed $q(z)$. The black dashed line shows $q(z)$ corresponding to the 
			$\Lambda$CDM model with $\Omega_{m,0} = 0.3$. A comparison among the four cases is shown in the extreme right column.}} \label{qplot_H0}
\end{figure*}

\begin{figure*}[h!]
	\begin{center}
		\includegraphics[angle=0, width=\textwidth]{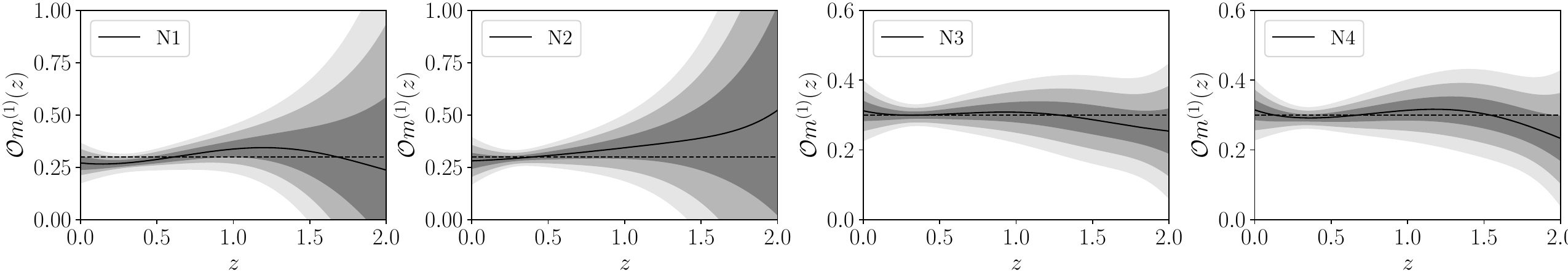}
	\end{center}
	\caption{{\small Plots for the ${\mathcal{O}m}^{(1)}$ diagnostics reconstructed from the combined datasets N1, N2, N3, N4 for a spatially flat universe ($\Omega_{k,0}=0$). 
			The solid black line represents the mean values of the reconstructed ${\mathcal{O}m}^{(1)}$. The black dashed line shows ${\mathcal{O}m}^{(1)}$ = $\Omega_{m,0}$ 
			corresponding to the $\Lambda$CDM model with $\Omega_{m,0} = 0.3$.}} \label{Omz_plot}
\end{figure*}

\subsection{Reconstruction of $q(z)$} \label{sec-5.3}

Finally, we plot the cosmological deceleration parameter $q(z)$ using the reconstructed values of the comoving distance $D(z)$, its derivatives $D'(z)$ and $D''(z)$, 
at different $z$ according to equation \eqref{qnew}. In Fig. \ref{qplot} and \ref{qplot_ok}, we plot the reconstructed $q(z)$ within 3$\sigma$ uncertainty regions 
for the combined datasets N1, N2, N3 and N4 considering two prior choices on the spatial curvature with $\Omega_{k,0} = 0$ and $\Omega_{k,0} = -0.0106 \pm 0.0065$ from 
Planck 2020 (TT, TE, EE+lowE+lensing) \cite{planck2020} respectively. The black solid lines represent the mean values and the shaded regions correspond to the 68\%, 
95\% and 99.7\% confidence levels for the reconstructed $q$. The black dashed line shows the evolution of $q(z)$ assuming the $\Lambda$CDM model with $\Omega_{m,0}=0.3$. 
The expected value of $q_{\Lambda\mbox{\tiny CDM}}$ at the present epoch is given by $q_{0_{\Lambda\mbox{\tiny CDM}}} = \frac{3}{2}\Omega_{m,0} - 1 = -0.55$.

For simplicity, we have assumed $\Lambda$CDM with $\Omega_{m,0} = 0.3$ as a reference model to compare our results. The chosen value $\Omega_{m,0} = 0.3$ is of 
course an approximation, but one can consider it to be approximately valid, since current observations from Planck 2020 probe \cite{planck2020}, SN-Ia Pantheon 
compilation \cite{pan1} and Dark Energy Survey \cite{abbott_des} do not predict any large deviations from this quoted value. In case we consider a different 
value of $\Omega_{m,0} = 0.3111 \pm 0.0056$ \cite{planck2020}, the expected value will be $q_{0_{\Lambda\mbox{\tiny CDM}}} = -0.5335 \pm 0.0084$ which is 
still well accommodated at the 1$\sigma$ confidence level of the reconstructed $q$ at the present epoch, $q_0$.

The mean values along with the associated 1$\sigma$ uncertainties of the reconstructed $q_0$, corresponding to the datasets N1, N2, N3 and N4 are shown in Table 
\ref{q0_tab}. An estimate for the late-time transition redshift $z_t$ where the reconstructed $q(z)$ shows a signature flip is also provided. This $z_t$ indicates 
the epoch when the expansion of the universe goes from a decelerating to an accelerating phase in the recent past.

\subsection{Effect of $H_0$ priors} \label{sec-5.4}

We further examine if the two different strategies for determining value of $H_0$ already mentioned in Sec. \ref{H0-priors} have any significant impact on 
the reconstruction of $q(z)$. We proceed with the analysis following a similar methodology as discussed in Sec. \ref{sec5.1}, \ref{sec-5.2} and finally Sec. 
\ref{sec-5.3} the only exception being that we have added the P20 or R21 $H_0$ estimates to the CC $H(z)$ dataset in the beginning. Finally, we reconstruct 
$q(z)$ for the following combinations, 

\begin{itemize}
	\item Set P1 - P20+CCB+SN, \item Set R1 - R21+CCB+SN, \item Set P2 - P20+CCM+SN, \item Set R2 - R21+CCM+SN,
	\item Set P3 - P20+CCB+SN+BAO, \item Set R3 - R21+CCB+SN+BAO, \item Set P4 - P20+CCM+SN+BAO, \item Set R4 - R21+CCM+SN+BAO.
\end{itemize}

Plots for the reconstructed $q(z)$ using the combined datasets P1, R1, P2, R2, P3, R3, P4 and R4 along with their respective 1$\sigma$, 2$\sigma$ and 3$\sigma$ 
uncertainties are shown in Fig. \ref{qplot_H0}. It is seen that inclusion of the P20 or R21 $H 0$ measurements does not contribute to any significant difference 
on the reconstruction of $q(z)$ in terms of allowing the $\Lambda$CDM model at the 2$\sigma$ confidence level. In case of the R1 combination, the mean reconstructed 
$q(z)$ shows the presence of a negative dip close to $z \simeq 1.9$ indicating another stint of acceleration in the recent past. For the N1 and P1 combinations, the 
possibility of this negative dip in $q$ can be perceived at higher redshift values exceeding the domain of reconstruction. However, this behaviour may not be 
statistically too significant as a positive $q$ is comfortably included at the $1\sigma$ confidence level.

\subsection{$\mathcal{O}m$ Diagnostics} \label{sec-5.5}

In context of the standard framework, we calculate the $\mathcal{O}m(z)$ diagnostic \cite{om1,om2,om3} which is given by
\begin{equation}
	\mathcal{O}m(z) = \frac{E^2(z)-1}{(1+z)^3 - 1} .  \label{Omz}
\end{equation} 

At the present epoch $z= 0$, the quantity $\mathcal{O}m(0)$ takes an indeterminate form. So, we obtain the modified $\mathcal{O}m(z)$ diagnostic function as

\begin{equation}
	{\mathcal{O}m}^{(1)}(z) = \frac{2 E(z) E'(z)}{3(1+z)^2 } .  \label{Omz_new}
\end{equation}

For a universe with an underlying expansion history $E(z)$ given by the $\Lambda$CDM model, $\mathcal{O}m(z)$ and also ${\mathcal{O}m}^{(1)}(z)$, will essentially 
be a constant, exactly equal to the magnitude of the matter density parameter at the present epoch $\Omega_{m,0}$. Therefore, any possible deviation from $\Omega_{m,0}$ 
can be used to draw inference on the dynamic nature of the universe.

Plots for the ${\mathcal{O}m}^{(1)}$ diagnostics from the combined N1, N2, N3 and N4 datasets are shown in Fig. \ref{Omz_plot} using the reconstructed $D(z)$ and $D'(z)$ 
in Eq. \eqref{Omz} corresponding to a spatially flat universe.  We observe that the $\Lambda$CDM model with a constant value of $\Omega_{m,0} = 0.3$ are consistent with 
the ${\mathcal{O}m}^{(1)}(z)$ reconstruction at the 2$\sigma$ confidence level.

\section{Reconstruction from Perturbation data}

For a universe composed of matter and dark energy in the background, the evolution of matter density contrast $\delta$ is given by

\begin{equation} \label{delta_def}
\delta = \frac{\delta \rho_m}{\rho_m}.
\end{equation}

This $\delta$, in a linearized approximation, obeys the following second order differential equation 

\begin{equation} \label{perturb_eqn}
\ddot{\delta}+ 2H \dot{\delta} - 4\pi G \rho_m \delta = 0.
\end{equation} 

Here, $\rho_m$ is the background matter density, $\delta \rho_m$ represents the first-order matter perturbation. On rewriting Eq. \eqref{perturb_eqn} as a function 
of the redshift $z$, the reduced Hubble parameter $E(z)$ can be expressed as an integral over the perturbation $\delta$ and its derivative \cite{zhang_li} as

\begin{equation} \label{E2_rsd}
\small E^2(z) = \frac{(1+z)^2}{\delta ' (z)^2} \left[ \delta '(z=0) ^2  -  3 \Omega_{m,0} \int_0^{z} \frac{\delta}{1+z} (-\delta ') \dif z \right].
\end{equation}

\begin{figure}[t!]
	\begin{center}
		\includegraphics[angle=0, width=0.49\textwidth]{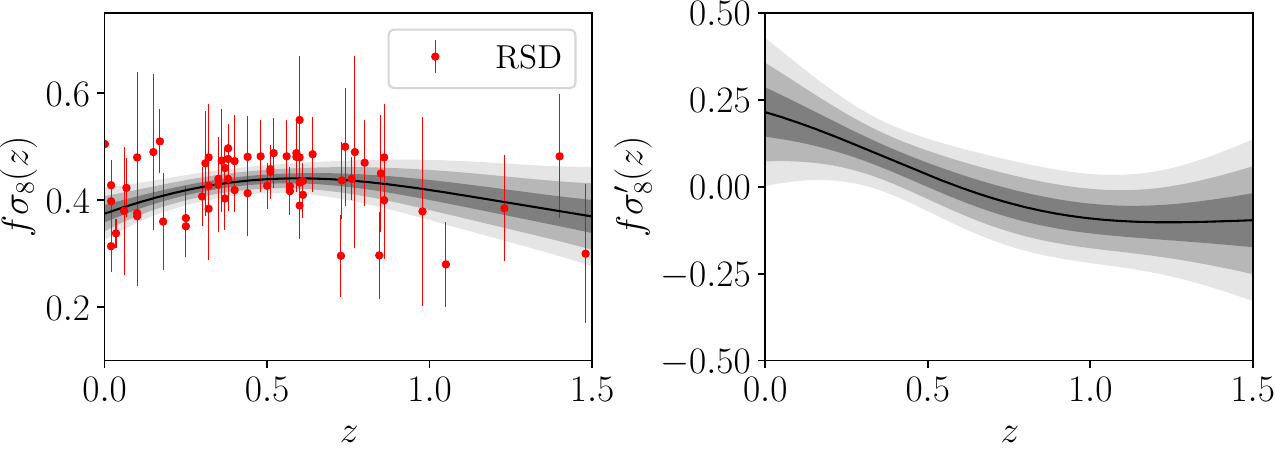}
	\end{center}
	\caption{{\small Plots for the GP reconstructed $f \sigma_8(z)$ and its derivative $\left[f \sigma_{8}\right]'$ from the RSD data. The solid black line represents the 
			mean values of the reconstructed functions. }} \label{fs8_plot}
\end{figure}

\begin{figure*}[h!]
	\begin{center}
		\includegraphics[angle=0, width=0.85\textwidth]{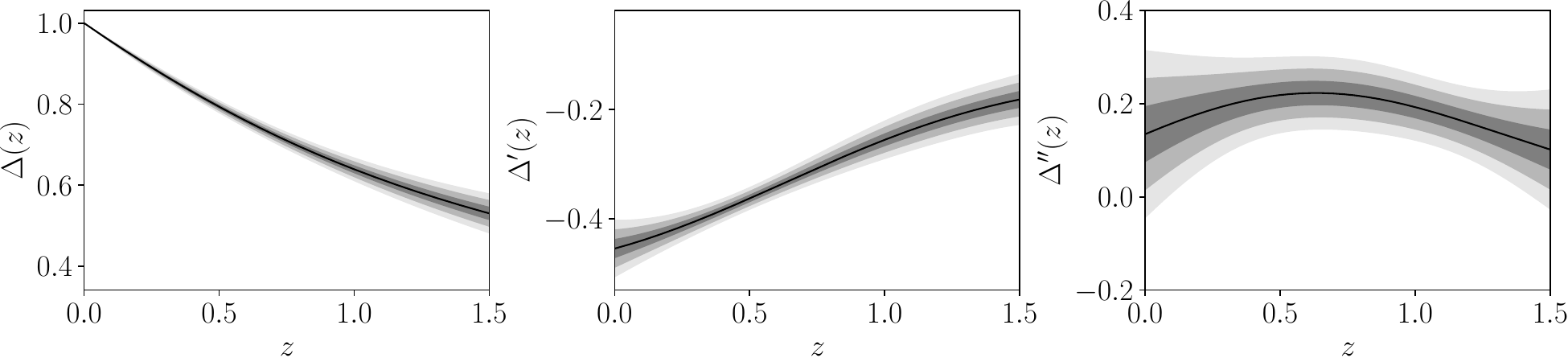}
	\end{center}
	\caption{{\small Plots for the reconstructed normalised perturbation $\Delta = \frac{\delta(z)}{\delta(z=0)}$ along with its higher derivatives $\Delta'(z)$ and 
			$\Delta''(z)$ from the RSD data. The solid black line represents the mean values of the reconstructed functions. }} \label{delta_plot}
\end{figure*}

\begin{figure*}[t!]
	\begin{center}
		\includegraphics[angle=0, width=0.9\textwidth]{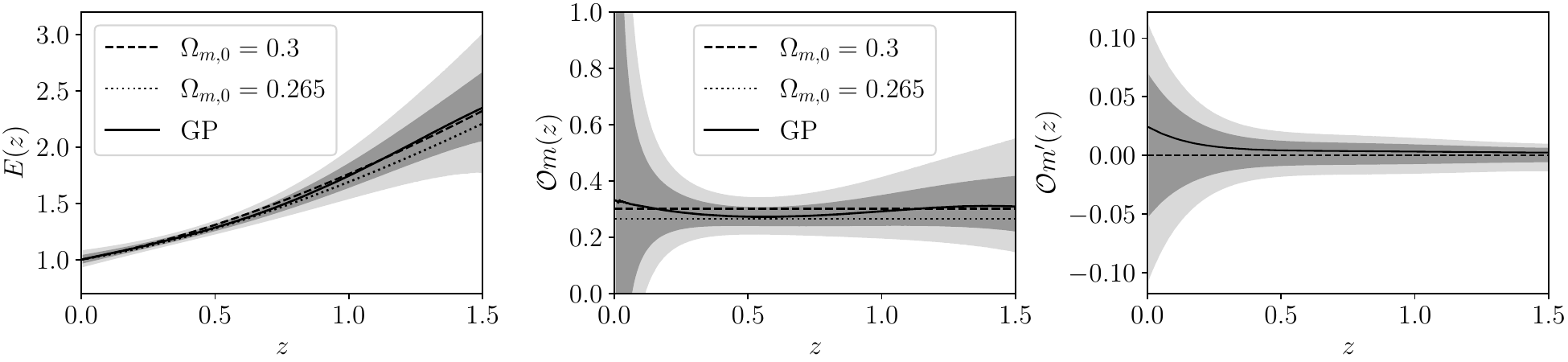}
	\end{center}
	\caption{{\small Plots for the reconstructed $E(z)$, $\mathcal{O}m(z)$ and $\mathcal{O}m'(z)$ diagnostics from the RSD data. The solid black lines represent the 
			mean values of the reconstructed functions. }} \label{EOm_rsd_plot}
\end{figure*}

Presently, cosmological observational surveys face the downside of not being able to provide a direct measurement of $\delta(z)$, but can successfully account 
for the related observations like $f\sigma_{8}$ from RSD. This $f \sigma_8$ is called the growth rate of structure, where $f$ is the growth rate, defined as 
the derivative of the logarithm of perturbation $\delta$ with respect to logarithm of the cosmic scale $a(t)$, i.e.

\begin{equation} \label{f_def}
f \equiv \frac{\dif \, \textmd{ln} \delta}{\dif \, \textmd{ln} a}  = -(1+z) \frac{\dif \, \textmd{ln} \delta}{\dif \, z} = -(1+z) \frac{\delta '}{\delta} .
\end{equation}

The function $\sigma_{8}$ is known as the linear theory root-mean-square mass fluctuation within a sphere of radius $8h^{-1}$ Mpc, with $h = \frac{H_0}
{100 \text{ km} \text{ Mpc}^{-1} \text{ s}^{-1}}$ being the dimensionless Hubble parameter at the present epoch, and is given by

\begin{equation} \label{s8_def}
\sigma_8 (z) = \sigma_8 (z=0) \frac{\delta (z)}{\delta (z=0)} .
\end{equation}

Therefore, the growth rate of structure can be consequently derived from Eq. \eqref{f_def} and \eqref{s8_def} as

\begin{equation}
	f \sigma_8 (z) = -\frac{\sigma_8 (z=0)}{\delta (z=0)} (1+z) \delta ' . \label{fs8}
\end{equation}

On integrating Eq. \eqref{fs8} followed by a some algebraic manipulation, we obtain

\begin{equation} \label{delta_end}
\delta  = \delta (z=0) - \frac{\delta (z=0)}{\sigma_8 (z=0)} \int_0^{z} \frac{f \sigma_8}{1+z} \dif z .
\end{equation}

For the reconstruction of $q(z)$ using the RSD data requires calculation of the integral 

\begin{equation}
	\mathcal{D} = \int_{0}^{z} \frac{f \sigma_8}{1+z} \dif z ,
\end{equation}

to obtain the perturbation $\delta$. Besides, the covariance uncertainty matrix associated with the $f \sigma_8$ RSD dataset need to be duly propagated into the 
uncertainty of $\delta(z)$. The statistical error associated with $E^2(z)$, defined in Eq. \eqref{E2_rsd}, can be expressed via the standard error propagation rule 
as 

\begin{equation}
	\sigma_{E^2}(z) = \left[\left( \frac{\partial E^2}{\partial \delta'} \right)^2 \sigma_{\delta'}^2 + \left( \frac{\partial E^2}{\partial \mathcal{D}} 
	\right)^2 \sigma_{\mathcal{D}}^2 \right] ^ \frac{1}{2}. 
\end{equation} 

Finally, we can reconstruct the deceleration parameter $q(z)$ using the expression 

\begin{equation}
	q(z) = -1 + \frac{1}{2}(1+z)\frac{\left[ E^2(z) \right]'}{E^2(z)}, \label{q_rsd_def} 
\end{equation}

where the uncertainty associated with $q(z)$ is propagated from the uncertainties in $E^2(z)$ and $\left[E^2(z)\right]'$ respectively. 

From close observation on Eq. \eqref{q_rsd_def} we infer that $q(z)$ is independent of the value of the perturbation $\delta$ at $z = 0$, but are directly 
dependent on the value of $\sigma_8$ and $\Omega_{m}$ at the present epoch, denoted as $\sigma_{8,0}$ and $\Omega_{m,0}$. For a self-consistent reconstruction 
of $q(z)$ from the RSD data, we need to provide the accurate values for $\sigma_{8,0}$ and $\Omega_{m,0}$. Instead of considering model-dependent estimates 
for $\sigma_{8,0}$ and $\Omega_{m,0}$, we attempt to constrain these parameters in a non-parametric way.

Although it is difficult to provide the analytical solution of Eq. \eqref{perturb_eqn}, assuming the universe to be spatially flat, an approximate solution is 
given in \cite{peebles,waga,fry,lightman,wang_stein,yggong} as

\begin{equation} \label{f_soln}
f(z) = \Omega_{m}^\gamma ,
\end{equation} 

where $\Omega_{m}(z) = \frac{\Omega_{m,0} (1+z)^3}{E^2(z)}$ and $\gamma$ is the growth index of perturbations corresponding to the background cosmological 
model. Therefore, $f \sigma_8$ in Eq. \eqref{fs8} is theoretically given by

\begin{equation} \label{fs8_final}
{f \sigma_8}^{\mbox{\tiny theo}} (z) = \sigma_{8,0} ~ \Omega_{m}^\gamma(z) \exp \left\lbrace \int_{0}^{z} - \frac{\Omega_{m}^\gamma(z')}{1+z'} \dif z'
\right\rbrace.
\end{equation}

\begin{figure*}[h!]
	\begin{center}
		\includegraphics[angle=0, width=0.9\textwidth]{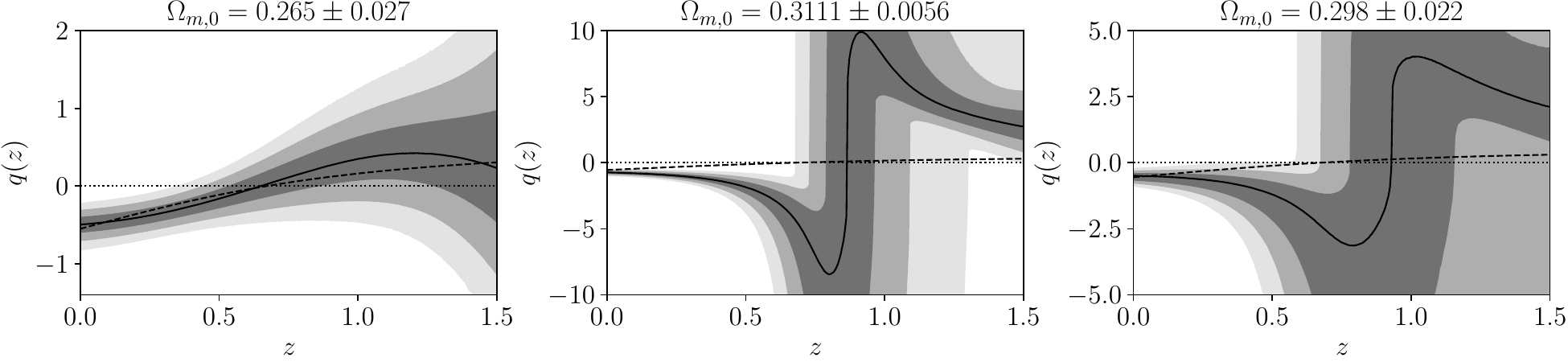}
	\end{center}
	\caption{{\small Plots for the deceleration parameter $q(z)$ reconstructed from the RSD dataset with different $\Omega_{m,0}$ priors. The solid black lines 
			represent the mean values of the reconstructed $q(z)$. The black dashed line shows $q(z)$ corresponding to the $\Lambda$CDM model with $\Omega_{m,0} 
			= 0.3$.}} \label{q_rsd_plot}
\end{figure*}

With a GP regression on the RSD data, we reconstruct the growth rate function $f\sigma_8(z)$ and its derivative $\left[f\sigma_{8}\right]'(z)$ at different redshift 
values and plot the results in Fig. \ref{fs8_plot}. At the present epoch, we obtain $f \sigma_{8} (z = 0) = 0.3748 \pm 0.0164$ and $\left[f \sigma_{8}\right]'(z = 0) 
= 0.2148 \pm 0.0709$ respectively. The marginalized constraints on $\Omega_{m,0}$ and $\gamma$ are obtained via a $\chi^2$ minimization between the theoretical 
$f \sigma_{8}^{\mbox{\tiny theo}}$ incorporating the reconstructed $E(z)$ from the combined CC and SN datasets in equation \eqref{fs8_final}, and the GP reconstructed 
$f\sigma_{8}^{\mbox{\tiny obs}}$ measurements from the RSD data, as
\begin{align}
\chi^2 &= \Delta  {\bf V^T} \textbf{Cov}^{-1} \Delta {\bf V}, \\
\Delta V_i &= f\sigma_{8}^{\mbox{\tiny obs}}(z_i) - f\sigma_{8}^{\mbox{\tiny theo}}(z_i) \\
\textbf{Cov} &= \textbf{Cov}^{\mbox{\tiny obs}} +\textbf{Cov}^{\mbox{\tiny theo}},
\label{chi2_rsd}
\end{align}

where $\textbf{Cov}^{\mbox{\tiny obs}}$ is the covariance matrix of $f\sigma_{8}^{\mbox{\tiny obs}}$ and $\textbf{Cov}^{\mbox{\tiny theo}}$ is the covariance matrix of 
the reconstructed $f\sigma_{8}^{\mbox{\tiny theo}}(z)$ which is defined in Eq. \eqref{fs8_final}. The parameter $\sigma_{8,0}$ serves as an additional constraint which 
can be eliminated by substituting $z = 0$ in Eq. \eqref{fs8_final}, such that 

\begin{equation}
\sigma_{8,0} = \frac{f \sigma_{8} (0)}{\Omega_{m,0}^\gamma} .  \label{s80-constraint}
\end{equation}

Adopting a Markov Chain Monte Carlo analysis with the assumption of uniform priors for $\Omega_{m,0} \in [0,1]$ and $\gamma \in [0.4,1.6]$, we obtain the best fit 
constraints as $\Omega_{m,0} = 0.265 \pm 0.027$ and $\gamma = 0.573 \pm 0.024$ respectively. The value of $\sigma_{8,0}$ is estimated from Eq. \eqref{s80-constraint} 
as $\sigma_{8,0} = 0.802 \pm 0.064$. With these parameter values we plot $\Delta(z)$, $\Delta'(z)$ and $\Delta''(z)$ in Fig. \ref{delta_plot} where $\Delta = \frac{
\delta(z)}{\delta(z=0)}$ is the normalised matter perturbation.

Finally, using the reconstructed $\Delta(z)$, $\Delta'(z)$ and $\Delta''(z)$ we plot the reduced Hubble parameter $E(z)$ besides the $\mathcal{O}m(z)$ and 
${\mathcal{O}m}^{\prime}(z)$ diagnostics in Fig. \ref{EOm_rsd_plot}. Here ${\mathcal{O}m}^{\prime}$ is recognized as the first order derivative of $\mathcal{O}m$ 
with respect to redshift $z$, which provides extra information regarding the possible variations in $\mathcal{O}m(z)$. Also, $\mathcal{O}m'$ utilizes the 
information from both $E(z)$ and $E'(z)$ reconstructions similar to the ${\mathcal{O}m}^{(1)}$ diagnostics, which includes additional information inferred 
from the GP analysis.

The plot for deceleration parameter $q(z)$ from the RSD data reconstructed utilizing Eq. \eqref{q_rsd_def} is shown in the extreme left of Fig. \ref{q_rsd_plot}. 
We observe that the deceleration parameter corresponding to the $\Lambda$CDM model is well contained at the 2$\sigma$ confidence level in the domain of reconstruction 
$0<z<1.5$. The reconstructed values of the deceleration parameter at the present epoch $q_0 = q(z=0)$ and the transition redshift $z_t$ are $q_0 = -0.496^{+0.098}_{-0.102} 
$ and $z_t = 0.651^{+0.213}_{-0.121}$ respectively.

Lastly, to test the influence of $\Omega_{m,0}$ on the reconstruction, we consider two cases namely $\Omega_{m,0} = 0.3111 \pm 0.0056$ from the Planck 2020 
\cite{planck2020} probe and $\Omega_{m,0} = 0.298 \pm 0.022 0$ from the Pantheon SN-Ia \cite{pan1} sample, as priors. For these cases, the parameter $\sigma_{8,0}$, 
is considered as $\sigma_{8,0} = 0.8102 \pm 0.0060$ from Planck 2020 \cite{planck2020}. We proceed with the GP reconstruction of $\Delta(z)$, $\Delta'(z)$ and $\Delta''
(z)$ to obtain the cosmic deceleration parameter $q(z)$, arising from these two cases. The results are shown in the centre and right columns of Fig. \ref{q_rsd_plot}. 
From the comparison in Fig. \ref{q_rsd_plot}, we find that the RSD data are highly sensitive to value of the matter density parameter $\Omega_{m,0}$ which leads to 
contrasting evolutionary scenarios for the reconstructed $q(z)$. In the first column, reconstruction with the RSD data for $\Omega_{m,0} = 0.265 \pm 0.027$ present an 
entirely different $q(z)$, when compared with the other two reconstructions shown in second and third columns of Fig. \ref{q_rsd_plot}. 

The plots shown in the central and right panels of Fig. \ref{q_rsd_plot} indicate a drastic change from a decelerated to an accelerated expansion of the universe close 
to $z \sim 0.8$, which is quite far from the transition redshift $z_t$ estimated from the combined background data. On the other hand, the left panel shows a more sedate 
transition at $z \sim 0.65$, much closer to the value of $z_t$ from the combined background data. Thus, if the value of $z_t$ is more trusted, we find that the value of 
${\Omega}_{m,0}$ is definitely less than $0.3$. This opens up a new possibility, the RSD data can help in constraining ${\Omega}_{m,0}$ and the value of $z_t$ can itself 
be observationally used as a new discriminator for cosmological models \cite{lima_zt}.

\section{Discussion}

The aim of this work is to reconstruct the deceleration parameter $q$ from recent observational data without any parametrization ansatz. As mentioned in the 
introduction, there are already quite a few efforts in this direction. However, as new data are pouring in and new techniques are evolving, revisiting the 
nature of $q$ with newer datasets is quite imperative. The present work is an endeavour towards that. We focus on a better model-independent treatment of the 
SN and BAO data, inclusion of all the recently updated systematic uncertainties in the CC data, as well as the RSD data. Reconstruction with the RSD data is 
a new feature that has been included in the present work.

We have utilized different combinations of the Pantheon compilation for Supernova distance modulus data, Cosmic Chronometer Hubble data with the full covariance 
matrix and Baryon Acoustic Oscillation data as mentioned in Sect. \ref{obs-data}. The GP reconstruction has been done considering the Mat\'{e}rn $9/2$ covariance 
function assuming a constant zero mean as prior. The effect of a non-zero spatial curvature from the Planck 2020 estimate, has been studied. The conflicting $H_0$ 
measurements from the recent Planck 2020 and Riess 2021 probes with a maximum discrepancy at $4.2\sigma$, has been analysed. In all cases studied, the common 
feature is that the mean curve for the reconstructed $q$ shows that the present acceleration has set in quite recently, for $z>0.5$ but well below $z=1$.

The deceleration parameter has also been reconstructed using the growth rate data from the RSD to investigate the effect of matter perturbations. We see 
that the value of $\delta$ at $z=0$ has no effect on the reconstruction of $q(z)$. We find that the matter density parameter $\Omega_{m,0}$ has a noticeable 
influence on the reconstruction of $q(z)$ as shown in Fig. \ref{q_rsd_plot}. The evolution of $q(z)$ obtained utilizing the RSD data is similar to the results 
from the combined CC and SN datasets in case of $\Omega_{m,0} = 0.265 \pm 0.027$, which is much lower than that of the Planck 2020 estimate. Therefore, the 
perturbation data have a more promising potential to distinguish between various dark energy models with distinct evolutionary scenarios.

We have further shown the reconstructed results for $\mathcal{O}m$ diagnostics in Fig. \ref{Omz_plot} and \ref{EOm_rsd_plot} from the background and growth rate 
data respectively. The plots reveal that the $\Lambda$CDM model is consistent within the redshift range of reconstruction at the 2$\sigma$ level. It is observed 
that the use of distance data from the Pantheon SN compilation leads to larger error bars at higher redshift, which is a contrasting feature when compared with the 
results obtained in \cite{haridasu} but similar to that of \cite{lin}.

The existing literature on the non-parametric reconstruction of $q$ indicates the presence of a dip in $q$ in the recent past. Bilicki \& Seikel \cite{bilicki} 
worked with either SN data (Union 2.1) or CC and BAO $H(z)$ data. Zhang and Xia \cite{xia} found that with the SN Union 2.1 or Union 2 data, a negative $q$ beyond 
a short-lived deceleration is not allowed in 2$\sigma$, but all the other data sets like CC, BAO  $H(z)$ and Gamma Ray Bursts (GRBs) indicate a dip in $q$ towards 
a negative value. Jesus, Valentim, Escobal \& Pereira \cite{jesus_nonpara} found constraints on the transition redshift $z_t$, along with a reconstruction of $q$ 
in a similar non-parametric Gaussian Process framework with CC and Pantheon SN data individually. A combination of all the data sets was commonly avoided in 
\cite{bilicki,xia,jesus_nonpara}. Lin, Li \& Tang \cite{lin} worked with the squared exponential covariance using a combination of the Pantheon SN and CC Hubble 
data. The authors in \cite{lin} found a dip in the best fit of reconstructed $q$, indicating an accelerated expansion in the recent past before a short-lived 
decelerated phase. Recently, G\'{o}mez-Valent \cite{adria} and Haridasu et al. \cite{haridasu} carried out two extensive analysis for the reconstruction of $q(z)$ 
using different combination of datasets. The Pantheon Supernova compilation of CANDELS and CLASH Multi-Cycle Treasury programs obtained by the HST (hereafter 
referred to as Pantheon + MCT) \cite{mct}, recent CC and BAO measurements, and the local $H_0$ measurement presented by HST photometry of long-period Milky Way 
Cepheid and GAIA parallaxes (hereafter referred to as R19) \cite{riess2019} was considered for the reconstruction of $q$ in \cite{haridasu, adria}. The authors 
in \cite{haridasu} employed a multi-task GP regression and found no dip in the best fit values of the reconstructed $q$. But such a dip, indicating an accelerated 
expansion in the recent past before a short-lived decelerated phase is very much allowed at the 1$\sigma$ confidence level. With the R19 data included, the 
presence of this dip in $q$ is quite clear in their work.

Our results are similar to those obtained in \cite{haridasu}, the difference being we have taken the Pantheon distance modulus compilation instead of the Pantheon 
+ MCT $E(z)$ measurements. It deserves mention that the supernova Pantheon + MCT $E(z)$ compilation are necessarily obtained assuming a spatially flat universe 
($\Omega_{k,0} = 0$). This pre-defined choice for a zero curvature could be a strong assumption that dampens the spirit of a model-independent analysis. As discussed 
in sections \ref{sec5.1} and \ref{sec-5.2}, the present work tests the possible effects of spatial curvature which is mostly absent in literature mentioned, except 
in the work of Zhang and Xia \cite{xia}, which however, ignores the combination of datasets. Comparison shows that the spatial curvature produces slight influence 
on the reconstruction. There is hardly any significant difference between the reconstructed values of the deceleration parameter.

A noticeable contrast can be found for the mean reconstructed function $q(z)$ in our analysis when compared to the results obtained in \cite{lin}, where the same 
combination of datasets (CC and Pantheon) were used as training data for GP regression. The difference lies in the methodology followed. We have opted for a better 
model-independent treatment of the Pantheon data i.e. estimating the marginalized constraints on the absolute magnitude $M_B$ instead of fixing it to the best-fitting 
$\Lambda$CDM value, as done in \cite{lin}. Moreover, our analysis accounts for all systematics within the CC data. Similarly, we have obtained the marginalized 
constraints on $r_d$ so as to eliminate the effect of any fiducial model-dependence linked with BAO measurements. For a reconstruction with individual sets, the results 
are independent of the choice on ${M_B}$ or $r_d$ because $q$ is a dimensionless quantity \cite{bilicki,xia, jesus_nonpara}. However, any arbitrary choice on these 
parameters can lead to inconsistencies in the results when working with various combinations of datasets. As a general note, we can comment that a fine tuning 
of these nuisance parameters, like $M_B$ and $r_d$, is desirable for a self-consistent combined analysis with high redshift future observations.

The mean values for the reconstructed $q_0$ and the late-time transition redshift $z_t$ obtained are given in Table \ref{q0_tab}. We have repeated the same analysis 
with the squared exponential covariance function and got closely similar results, for example allowing the $\Lambda$CDM model at the 2$\sigma$ level. We find that 
this agreement with $\Lambda$CDM  strongly depends on the redshift of reconstruction and is much better at the low redshift regime. At higher $z$, the mean 
reconstructed curve deviates from the $\Lambda$CDM behaviour with large error bounds. The two competing values of $H_0$, namely the P20 and R21, can hardly make 
any qualitative difference in the results as shown in Fig. \ref{qplot_H0}, except for the R1 combination where the mean values of the reconstructed $q(z)$ shows a 
negative dip. The N1 and P1 combinations show the possibility of this negative dip in $q$ at higher redshift values. However, this negative dip at high $z$ does 
not seem to have any high statistical significance, as the reconstructed $q$ in the recent past allows a decelerated expansion as well at the 1$\sigma$ confidence 
level for $z>z_t$. It should be emphasized that from $z=0$ to roughly $z=0.5$, no deceleration is allowed even in 3$\sigma$ for all the cases. We found that the 
reconstructed $q(z)$ shows an approximately linear behaviour in $z$ for the redshift range $0<z<z_t$, which closely mimics the $\Lambda$CDM behaviour. At higher 
redshift, beyond $z>1$, the reconstructed $q$ shows a non-monotonic behaviour for the combined CC and Pantheon datasets. Inclusion of the BAO data gives rise to 
an oscillatory behaviour in the reconstructed $q$. 
The presence of large uncertainties in the reconstructed $q$ at higher redshift arises to lesser availability of observational data at high $z$.

In conclusion, we can say that not only the nature of dark energy but also the evolution history of the universe is yet to be properly ascertained. The number 
density of the CC, SN, BAO and RSD data is well concentrated up to $z\simeq1$ (as shown in Fig. \ref{D_N1}, \ref{D_N2}, \ref{D_N3}, \ref{D_N4}, and \ref{fs8_plot}). 
The availability of data in the redshift range $1<z<2$, is much lower and this can have a considerable effect on the reconstruction, such as larger uncertainties 
in the reconstructed function particularly with increasing redshift. We agree with Lin, Li \& Tang \cite{lin} that we need more data and also perhaps a better 
model-independent treatment of the data as well. The reconstruction of kinematic parameters like $q$ will have to be renewed time and again with newer data in 
search of a better understanding of the evolution.

\section*{Acknowledgements}

The authors would like to sincerely thank the anonymous referee for his constructive suggestions and valuable comments that led to a substantial improvement 
of this work. The authors also thank Ankan Mukherjee for useful discussions and suggestions.

\section*{Data Availability}

The authors confirm that all relevant data, included in the manuscript, are either from public domain or from published papers, all of which are duly cited. 

\section*{References}


\end{document}